\documentclass{emulateapj}\usepackage{apjfonts}

\newcommand{\afuv}{\mbox{$A_{\rm FUV}$}}
\newcommand{\aha}{\mbox{$A_{\rm H\alpha}$}}
\newcommand{\as}{\mbox{$''$}}
\newcommand{\dg}{\mbox{$^{\circ}$}}

\newcommand{\ewha}{\mbox{$EW(\Halpha)$}}
\newcommand{\fesc}{\mbox{$f_{\rm esc}$}}
\newcommand{\ffuv}{\mbox{$f_{\rm FUV}$}}
\newcommand{\fr}{\mbox{$f_{\rm R}$}}
\newcommand{\fion}[2]{\mbox{\rm [{#1}\thinspace{\footnotesize {#2}}]}} 
\newcommand{\Fha}{\mbox{$F_{\rm H\alpha}$}}
\newcommand{\Fhafuv}{\mbox{$F_{\rm H\alpha}/f_{\rm FUV}$}}
\newcommand{\fFhafuv}{\mbox{$\frac{F_{\rm H\alpha}}{f_{\rm FUV}}$}}
\newcommand{\Fhafuvr}{\mbox{$F_{\rm H\alpha}'/f_{\rm FUV}'$}}
\newcommand{\fFhafuvr}{\mbox{$\frac{F_{\rm H\alpha}'}{f_{\rm FUV}'}$}}
\newcommand{\Fhafir}{\mbox{$F_{\rm H\alpha}/F_{\rm FIR}$}}
\newcommand{\HI}{\mbox {\sc H i}}
\newcommand{\HII}{\mbox {\sc H ii}}
\newcommand{\HIPASS}{{\sc HiPASS}}
\newcommand{\Hline}[1]{\mbox{H{\footnotesize {#1}}}}
\newcommand{\Halpha}{\Hline{\mbox{$\alpha$}}}

\newcommand{\LHa}{\mbox{$L_{\footnotesize  H{\tiny \alpha}}$}}
\newcommand{\LR}{\mbox{$L_R$}}
\newcommand{\Mclus}{\mbox{${\cal M}_{\rm cluster}$}}

\newcommand{\MHI}{\mbox{${\cal M}_{\rm HI}$}}
\newcommand{\Mstar}{\mbox{${\cal M}_\star$}}
\newcommand{\Mstaravg}{\mbox{$\langle{\cal M}_\star\rangle$}}
\newcommand{\Msun}{\mbox{${\cal M}_\odot$}}
\newcommand{\Mlow}{\mbox{${\cal M}_l$}}

\newcommand{\Mupp}{\mbox{${\cal M}_u$}}
\newcommand{\Most}{\mbox{${\cal M}_{\rm O}$}}
\newcommand{\Pext}{\mbox{$P_{\rm mid}$}}
\newcommand{\rxy}{\mbox{$r_{\rm xy}$}}
\newcommand{\SHa}{\mbox{$\Sigma_{\rm H\alpha}$}}
\newcommand{\SHar}{\mbox{$\Sigma_{\rm H\alpha}'$}}
\newcommand{\SSFR}{\mbox{$\Sigma_{\rm SFR}$}}
\newcommand{\SR}{\mbox{$\Sigma_R$}}
\newcommand{\SRr}{\mbox{$\Sigma_R'$}}
\newcommand{\tdyn}{\mbox{$\tau_{\rm dyn}$}}
\newcommand{\torb}{\mbox{$\tau_{\rm orb}$}}
\newcommand{\Vrot}{\mbox{$V_{\rm rot}$}}
\newcommand{\na}{{New A}}%

\shorttitle{Non-Uniform IMF}
\shortauthors{Meurer et al.}
\slugcomment{ApJ in press}

\begin{document}

\title{Evidence for a Non-Uniform Initial Mass Function in the Local
  Universe$^*$}

\author{Gerhardt R. Meurer\altaffilmark{1}, O.I.\
  Wong\altaffilmark{2,3}, J.H.\ Kim\altaffilmark{1}, D.J.\
  Hanish\altaffilmark{1}, T.M.\ Heckman\altaffilmark{1}, J.\
  Werk\altaffilmark{4}, J.\ Bland-Hawthorn\altaffilmark{5}, M.A.\
  Dopita\altaffilmark{6}, M.A.\ Zwaan\altaffilmark{7}, B.\
  Koribalski\altaffilmark{8}, M.\ Seibert\altaffilmark{9}, D.A.\
  Thilker\altaffilmark{1}, H.C.\ Ferguson\altaffilmark{10}, R.L.\
  Webster\altaffilmark{3}, M.E.\ Putman\altaffilmark{11}, P.M.\
  Knezek\altaffilmark{12}, M.T.\ Doyle\altaffilmark{13}, M.J.\
  Drinkwater\altaffilmark{13}, C.G.\ Hoopes\altaffilmark{1}, V.A.\
  Kilborn\altaffilmark{14}, M.\ Meyer\altaffilmark{15}, E.V.\
  Ryan-Weber\altaffilmark{16}, R.C.\ Smith\altaffilmark{17}, and L.\
  Staveley-Smith\altaffilmark{15}}

\altaffiltext{*}{Based on observations made with the NASA Galaxy
                 Evolution Explorer.  GALEX is operated for NASA by the
                 California Institute of Technology under NASA contract
                 NAS5-98034.}
\altaffiltext{1}{Department of Physics and Astronomy, The Johns Hopkins
                 University, 3400 North Charles Street, Baltimore, MD 
		 21218; meurer@pha.jhu.edu}
\altaffiltext{2}{Department of Astronomy, Yale University, New Haven, 
                 CT 06520}
\altaffiltext{3}{School of Physics, University of Melbourne, VIC 3010, 
                 Australia} 
\altaffiltext{4}{University of Michigan, Department of Astronomy, 830
                 Denison Building, Ann Arbor, MI 48109-1042} 
\altaffiltext{5}{Institute of Astronomy, School of Physics, University
                 of Sydney, Australia}
\altaffiltext{6}{Research School of Astronomy and Astrophysics, 
                 Australian National University, Cotter Road, Weston 
                 Creek, ACT 2611, Australia}
\altaffiltext{7}{European Southern Observatory,
                 Karl-Schwarzschild-Str.\ 2, 85748 Garching b.\ 
                 M\"{u}nchen, Germany}
\altaffiltext{8}{Australia Telescope National Facility, CSIRO, P.O. Box
                 76, Epping, NSW 1710, Australia}
\altaffiltext{9}{Observatories of the Carnegie Institute of Washington,
                 Pasadena, CA, 91101}
\altaffiltext{10}{Space Telescope Science Institute, 3700 San Martin
                  Drive, Baltimore, MD 21218}
\altaffiltext{11}{Department of Astronomy, Columbia University, 550 West
                  120th Street, New York, NY 10027}
\altaffiltext{12}{WIYN Consortium, Inc., 950 North Cherry Avenue, Tucson,
                  AZ 85726}
\altaffiltext{13}{Department of Physics, University of Queensland, 
                  Brisbane, QLD 4072, Australia}
\altaffiltext{14}{Centre for Astrophysics and Supercomputing, Swinburne 
                  University of Technology, Mail 31, PO Box 218, 
                  Hawthorn, VIC 3122, Australia}
\altaffiltext{15}{School of Physics, University of Western Australia,
                  Crawley, WA 6009, Australia }
\altaffiltext{16}{Institute of Astronomy, Madingley Road, Cambridge 
                  CB3 0HA, United Kingdom}
\altaffiltext{17}{Cerro Tololo Inter-American Observatory (CTIO), Casilla
                  603, La Serena, Chile} 

\begin{abstract}
Many of the results in modern astrophysics rest on the notion that the
Initial Mass Function (IMF) is universal.  Our observations of a
sample of \HI\ selected galaxies in the light of \Halpha\ and the
far-ultraviolet (FUV) challenge this result.  The extinction corrected
flux ratio \Fhafuv\ from these two tracers of star formation shows
strong correlations with the surface-brightness in \Halpha\ and the
$R$ band: Low Surface Brightness (LSB) galaxies have lower \Fhafuv\
ratios compared to High Surface Brightness (HSB) galaxies as well as
compared to expectations from equilibrium models of constant star
formation rate (SFR) using commonly favored IMF parameters.  Weaker
but significant correlations of \Fhafuv\ with luminosity, rotational
velocity and dynamical mass are found as well as a systematic trend
with morphology.  The correlated variations of \Fhafuv\ with other
global parameters are thus part of the larger family of galaxy scaling
relations.  The \Fhafuv\ correlations can not be due to residual
extinction correction errors, while systematic variations in the star
formation history can not explain the trends with both \Halpha\ and
$R$ surface brightness nor with other global properties.  The
possibility that LSB galaxies have a higher escape fraction of
ionizing photons seems inconsistent with their high gas fraction, and
observations of color-magnitude diagrams of a few systems which
indicate a real deficit of O stars. The most plausible explanation for
the correlations is the systematic variations of the upper mass limit
\Mupp\ and/or the slope $\gamma$ which define the upper end of the
IMF.  We outline a scenario of pressure driving the correlations by
setting the efficiency of the formation of the dense star clusters
where the highest mass stars preferentially form.  Our results imply
that the star formation rate measured in a galaxy is highly sensitive
to the tracer used in the measurement.  A non-universal IMF would also
call into question interpretation of metal abundance patterns in dwarf
galaxies as well star formation histories derived from color magnitude
diagrams.
\end{abstract}


\section{Introduction}\label{s:intro}

A key concept in our understanding of the evolution of galaxies is the
Initial Mass Function (IMF), which gives the statistical distribution of
masses of stars that form in a single event.  Typically the IMF is
parametrized as a power law, or a series of broken power-laws, in
stellar mass, \Mstar, where the key parameters are the power-law index
$\gamma$ and the lower and upper mass limits to the stars formed \Mlow\
and \Mupp\ respectively.  The IMF is crucial in interpreting the nature
of galaxies because the light we observe is dominated by the highest
mass stars while the total mass in stars is dominated by the lower mass
stars.  If we know the IMF of a stellar population, and its age, then we
can estimate its mass from photometry.  Similarly when looking at
tracers of star formation, such as \Halpha\ or ultraviolet (UV)
emission, the IMF allows us to convert from luminosity to a star
formation rate (SFR).

The crucial assumption underpinning much of the use of the IMF is that
it is universal \citep{g01}; it does not vary within galaxies or between galaxies.
This certainly seems to be the case for star clusters
\citep{kroupa01,kroupa02}.  Since most star formation occurs in star
clusters \citep{ll03} a universal IMF seems plausible, and is well
accepted in the literature. 

In recent years there has been
accumulating evidence that the IMF is not universal 
\citep[see review of][]{e08a}.  This sets the stage for our study in
which we challenge the notion that the upper end of the IMF is uniform
using measurements of the integrated \Halpha\ and UV emission in a
sample of \HI\ selected galaxies from the Survey of Ionization in
Neutral Gas Galaxies (SINGG), and the Survey of Ultraviolet emission in
Neutral Gas Galaxies (SUNGG).  \Halpha\ emission traces the presence of
ionizing O stars, which have initial masses $\Mstar \gtrsim 20 \Msun$,
while UV emission traces both O and B stars with $\Mstar \gtrsim 3
\Msun$, giving the leverage in mass needed to probe the IMF.  

The \HI\
selection performed by SINGG and SUNGG results in a sample of star
forming galaxies spanning the full range of morphologies seen in star
forming galaxies with very few \Halpha\ non-detections
\citep[][hereafter M06]{singg1} and can fully account for the cosmic
density of star formation in the local universe \citep[][hereafter
H06]{singg2}.  By using this sample we avoid the issue of selection
effects biasing towards particular extreme star formation histories
(SFH) such as starburst galaxies.  While the SFH of the universe as
whole appears to have undergone a dramatic factor of 10 decline since $z
\sim 1$ ($\sim 8$ Gyr ago for $H_0 = 70 \, {\rm km\, s^{-1} Mpc^{-1}}$ 
which we adopt here) this is still long compared to the
lifetimes of the O ($\lesssim 7$ Myr) and B ($\lesssim 320$ Myr) stars
probed in this study, so the SFH should be effectively constant for the sample
as a whole.  

There are a few other advantages of our approach.  We use
integrated fluxes, so in
the parlance of \citet{wk05} we are dealing with the Integrated Galaxial
IMF (IGIMF). Unlike the case of color-magnitude diagram
analysis, crowding effects are irrelevant as are concerns as to the
birth place of stars and their eventual dispersal in the field.  
Stochastic effects are minimized by integrating over whole galaxies, and
the results are more comparable to studies of distant galaxies for which
little more than total fluxes are available.

This paper is organized as follows.  In Sec.~\ref{s:quant} we discuss the
sample selection, the data used, the measurements of these data which we
make, and the necessary corrections to the measurements.  In
Sec.~\ref{s:popmod} we demonstrate the sensitivity of the measured
fluxes to the IMF parameters and present a benchmark stellar population
model for interpreting the results.  Section~\ref{s:results} shows our
primary result - that there is a strong correlation of the \Halpha\ to
UV flux ratio with surface brightness of both \Halpha\ and $R$ band
emission, as well as weaker correlations of this ratio with other
quantities.  In Sec.~\ref{s:ratpars} we discuss other physical
parameters that can affect this ratio and argue that the IMF parameters
are most likely driving the observed correlations.  Section~\ref{s:disc}
places our results in the context other work, presents a physical
scenario for pressure driven IMF variations driving the observed
correlations, and discusses the implications of our results.
Section~\ref{s:summ} summarizes our results and suggests future
observations.

\section{Observations and measurements}\label{s:quant}

\begin{figure}
\epsscale{1.1}
\plotone{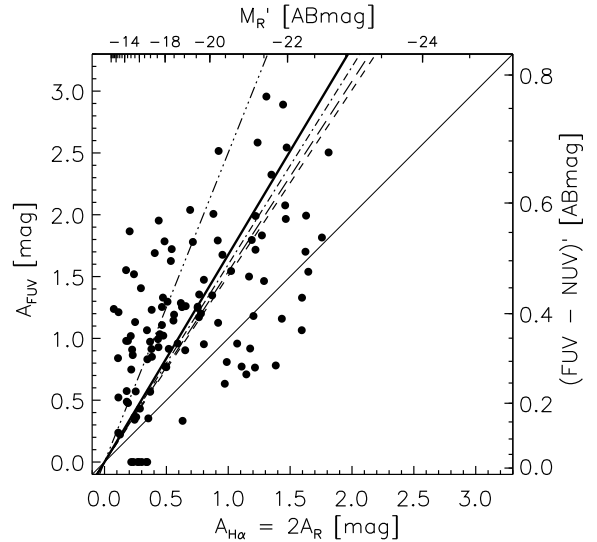}
\caption[]{The adopted \Halpha\ and FUV dust absorption
  in magnitudes, \aha\ and \afuv, respectively plotted against each
  other.  These are determined from the $R$ band absolute magnitude
  ($M_R'$) and the FUV to NUV color (FUV -- NUV)$'$ corrected for Galactic
  dust absorption but not internal dust absorption. These quantities are
  shown on the alternate axes (top and right, respectively).  The thin
  solid line shows the unity relation ($\afuv = \aha$).  The heavy solid
  line shows the average ratio for our sample.  The $\afuv/\aha$ ratio
  expected for the \citep{c00} and \citep{ccm89} dust absoprption laws are
  shown with the long dashed and dash - triple dot lines, respectively;
  while the the ratio $\afuv/(2\aha)$ is shown by the short dashed and
  dash-dot lines respectively. \label{f:avsa}}
\end{figure}

\subsection{Sample Selection}

Our sample consists of 103 galaxies with both \Halpha\ observations from
SINGG and UV observations from SUNGG.  The ultimate parent of both these
surveys is the \HI\ Parkes All Sky Survey, \HIPASS\
\citep{hipass1,bgc04}.  M06 discuss the selection of the SINGG sample in
detail.  In brief, the sample was selected uniformly in the log of \HI\
mass, \MHI, with the nearest galaxies preferred in order to yield the
best spatial resolution. Only \HI\ properties were used to select the
SINGG sample.  SUNGG is a Cycle 1 GALEX Legacy Survey, whose targets
were all chosen from the SINGG sample \citep{wongthesis,sunggatlas}.  
Because of the large GALEX field of view (1.1\dg
diameter), some SINGG galaxies not in the original SUNGG selection are
also included in the measurements presented here.  As shown by M06,
\HIPASS\ sources often correspond to multiple emission line galaxies
(ELGs) in our narrow-band images (M06).  By default we consider them part
of the \HIPASS\ source targeted.  However, we have excluded five
sources in multiple ELG fields from the analysis after inspecting the
optical and UV images.  Three of these are barely larger than the
optical PSF and may be background galaxies (e.g.\ \fion{O}{III} emitters
at $z = 0.3$), one may be part of a larger galaxy and not a galaxy in
its own right, and the last is not totally within the CCD frame.

\subsection{Optical Observations}

The \Halpha\ and $R$ band observations were obtained primarily with the
CTIO 1.5m telescope, while some targets were observed with the CTIO 0.9m
telescope to a similar depth, and at a similar plate scale.  M06
describes the method of data acquisition, reduction and analysis.  The
data we use here are measurements of integrated flux and surface
brightness.  As detailed in M06 these are measured using concentric
elliptical aperture photometry after masking out obvious foreground
stars, background galaxies and other image blemishes from the sky
subtracted images.  Plots of enclosed flux as a function of aperture
major axis radius are used to determine the total flux, $F$; the radii
enclosing 50\%\ and 90\%\ of the flux, $r_{50}$ and $r_{90}$; and the
effective surface brightness $\Sigma$ which is the face-on surface brightness
within $r_{50}$: $\Sigma = F/(2\pi r_{50}^2)$, where here we use $F$ to
denote fluxes, or count rates in general. Later we become more specific
and use $F$ to denote integrated line fluxes and $f$ to distinguish
continuum flux density.

The \Halpha\ signal to noise ratio is low, $S/N < 2$ within the
outermost measurement aperture for 19 of the ELGs.  In most of these
\Halpha\ is clearly detected, but its signal is washed out over the
large apertures needed to measure the total flux.  For these cases we
estimate the total \Fha\ as twice the \Halpha\ flux within $r_{50}$
determined from the $R$ band.  This correction factor is consistent with
the observation that the \Halpha\ and $R$ emission typically have equal
$r_{50}$ values (H06).  Since the measurement errors are much smaller
within $r_{50}$ this provides significant detections for 5 of the low
S/N sources.  We use upper limits based on the $r_{50}$ extrapolated
flux for the remaining 14 sources.  Of these, only one, HIPASS
J1321$-$31, has no \Halpha\ emission discernible in our images.  The
optical counterpart is a Low Surface Brightness (LSB) galaxy identified
using \HI\ synthesis imaging observations presented by Ryan-Weber et
al.\ (2003; also shown by Grossi et al.\ 2007)\nocite{rws03,grossi07}.
\citet{pritzl03} also note that no \Halpha\ is seen in their WIYN 3.5m
observations of J1321-31.  This source is discussed further in
Sec~\ref{s:stoc}. 

\subsection{Ultraviolet Observations}

The FUV and NUV GALEX images were typically 1500s (one eclipse) in
duration, the same depth as the GALEX Nearby Galaxy Atlas \citep{gdp07}.
We make measurements from the combined flux calibrated images of each
field taken from the GALEX pipeline \citep{morrissey07}. We perform our
own background subtraction since the angular diameter of our galaxies is
often $\gtrsim 3.2'$ which is the sky grid size used for GALEX pipeline
processing \citep{morrissey07}.  Fluxes, radii ($r_{50}$, $r_{90}$), and
surface brightness in the FUV and NUV are measured in a manner analogous
to that done for SINGG - using enclosed flux curves after masking out
foreground and background objects.  

The apertures used by SINGG and SUNGG are configured
independently. Hence the UV and optical apertures generally are not
matched, with the UV aperture almost always being larger.  Comparison of
the aperture areas with the $r_{50}$ and $r_{90}$ measurements, shows
that the smaller of the UV or optical aperture should recover the
majority of the flux at all wavelengths in almost all cases.
Specifically, if the \Halpha\ and UV light trace each other (as expected
for a universal IMF) the bias induced by mismatched apertures in
$\log(\Fhafuv)$ should be $\leq 0.05$ dex in 81\%\ of the sample, may be
as high as 0.3 dex in 17\%\ and may be higher in 2\%\ of the sample (2
cases).  The actual bias depends on how much light is outside the
optical measurement aperture, and may be considerably lower than these
estimates if the IMF at large radii is deficient in high mass stars as
found in some spiral galaxies \citep{thilker05,b07}.

\subsection{Error bars}

The error bars we display account for the measurement errors only.  The
errors in the optical quantities are discussed in detail in M06, and
include terms for the sky uncertainty and the continuum subtraction.
The errors in the UV quantities include terms for photon statistics and
the background uncertainty \citep{wongthesis,sunggatlas}.

\begin{figure}
\epsscale{1.1}
\plotone{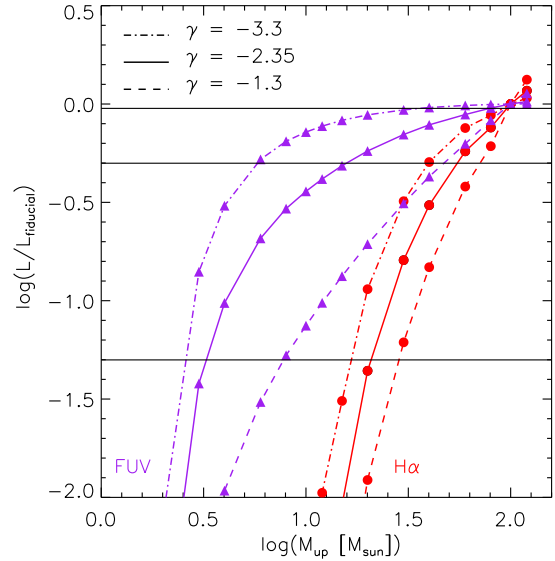}
\caption[]{Sequences of population synthesis models showing the effects
  on the luminosity in \Halpha\ and FUV of varying \Mupp\ at fixed
  $\gamma$, relative to a fiducial model having $\Mupp = 100\, \Msun$.
  Three sequnces corresponding to $\gamma = -3.3, -2.35, -1.3$
  (differentiated by line style as shown in the legend) are shown for each band with the
  \Halpha\ sequences shown as (red) connected dots and the FUV sequences
  as (purple) connected triangles. Horizontal lines from top to bottom
  indicate luminosities of 95\%, 50\%, and 5\% relative to the fiducial
  model.  See the text for further details of
  the models.  \label{f:mupmods}}
\end{figure}

\subsection{Corrections for Emission Line contamination and Galactic
  Dust Absorption}

A variety of corrections to the measurements are required to transform
them to intrinsic quantities.  M06 details the correction of
the \Halpha\ measurements for \fion{N}{II} emission and Balmer
absorption line contamination.  The \Halpha, $R$, FUV, NUV measurements
are all corrected for foreground Galactic dust absorption using the
reddening maps of \citet{sfd98} and the extinction law of \citet{ccm89}.

\subsection{Correction for Internal Dust Absorption}\label{s:cdust}

Internal dust absorption can have a substantial effect on incident
fluxes, especially in the UV,  causing systematic biases in
\Fhafuv.  We correct the optical and UV data using separate
empirical correlations.  

For the UV data we base the correction on a fit to the correlation
between the ratio of total far infrared (FIR) flux and the FUV flux and
the FUV - NUV color.  The total FIR flux was determined from IRAS
fluxes following \citet{dhcsk01} for 73 galaxies in the SUNGG sample
that were detected by IRAS. The IRX-$\beta$ data and fit for our sample
is shown in \citet{sunggatlas}.  Similar correlations are shown by
\citet{cbbgbgsmm06}, \citet{b07}, and \citet{gdp07}.  The fitted
relationship thus provides a method to recover the light absorbed by
dust and re-radiated in the far-infrared.

\begin{deluxetable*}{ccccccc}
  \tablecolumns{7}
  \tablewidth{0pt}
  \tablecaption{Contribution to luminosity by stellar mass\label{t:mupmods}}
  \tablehead{\colhead{contribution} &
             \multicolumn{3}{c}{H$\alpha$} &
             \multicolumn{3}{c}{FUV}\\
              \colhead{level} &
              \colhead{$\gamma = -3.3$} &
              \colhead{$\gamma = -2.35$} &
              \colhead{$\gamma = -1.3$} &
              \colhead{$\gamma = -3.3$} &
              \colhead{$\gamma = -2.35$} &
              \colhead{$\gamma = -1.3$}\\
              \colhead{(1)} &
              \colhead{(2)} &
              \colhead{(3)} &
              \colhead{(4)} &
              \colhead{(5)} &
              \colhead{(6)} &
              \colhead{(7)}}
\startdata
    5\%  &   17 &  23  & 28   &   2.5 &  3.3  & 7.7 \nl
   50\%  &   40 &  54  & 71   &   5.7 &  16   & 47 \nl
   95\%  &   91 &  95 &  97   &   35  &  78   & 94 \nl
\enddata
\tablecomments{The values in this table correspond to the intersection
  of the horizontal lines with the curves shown in Fig.~\ref{f:mupmods}.
  They show the contribution of different
  stellar masses to the luminosity of a solar metallicity stellar
  population that has been forming stars for 1 Gyr.  The stellar
  population has a single power law IMF with mass range of 0.1 to 100
  \Msun.  The table gives the upper end of the mass range, in \Msun,
  that contributes the first N\%\ (specified in column 1) to the
  luminosity in \Halpha\ (columns 2 - 4) and FUV (columns 5 - 7) for
  three different IMF slopes $\gamma$.  Thus for a Salpeter IMF ($\gamma
  = -2.35$) 5\%, 50\%, and 95\%\ of the \Halpha\ luminosity comes from
  stars having masses up to 23 \Msun, 54 \Msun, and 95 \Msun
  respectively, while 5\%, 50\%, and 95\%\ of the FUV luminosity comes
  from stars having masses up to 3.3 \Msun, 16 \Msun, and 78 \Msun\
  respectively. }
\end{deluxetable*}

The optical fluxes were corrected for internal dust using a relationship
between \Halpha\ dust attenuation derived from Balmer line ratios and
$R$ band absolute magnitude prior to internal dust correction, $M_R'$,
found and presented by \citet{hwbod04} using data from the Nearby Field
Galaxy Survey \citep{jansen00,jffc00}.  The $R$ band internal dust
attenuation, in magnitudes was taken to be half that of \Halpha\ to
account for ``differential'' dust absorption: emission lines are more
effected by dust than the stellar continuum \citep{fct88,c94},
presumably because high-mass star formation is more highly
correlated with dust than the general field star population.  M06
present a population synthesis model attenuated by dust using this
prescription in order to explain the relationship between 
\LR\ and the ratio of \Halpha\ to FIR fluxes,
\Fhafir. The model used had a solar
metallicity, a Salpeter IMF over the mass range of 0.1 to 100
\Msun, and  a constant SFR for a duration 100 Myr,
and was calculated using Starburst99 \citep{starburst99}, while the dust
followed the \citet{c00} dust attenuation model.  While this model
accounted for the shape of the \Fhafir\ versus \LR\ relationship,
there is an offset: \Fhafir\ is predicted to be too high by a factor of
$\sim 3$.  There are various possible explanations for this discrepancy,
some of which are discussed by M06.  In particular, they note that
adopting a steeper $\gamma$ and/or lower \Mupp\ will help alleviate the
discrepancy.  Hence, the low measured \Fhafir\ values may be related to
the low \Fhafuv\ values that are the primary concern of this
investigation. Other possible factors that could contribute to the M06
\Fhafir\ model being too high include: too short of a duration for star
formation, a metallicity that is too low, escaping ionizing photons,
choice of stellar population models, a
dust absorption model that is too differential, and an error in the FIR
bolometric correction.  It is beyond the scope of this paper to improve
the \Fhafir\ versus $M_R'$ model of M06.  However, except for the dust 
bolometric correction,  we do address all of
these issues here with regard to our
\Fhafuv\ results.  Since the overall shape of the $M_R'$
-- \Fhafir\ relationship agrees with the M06 model, then using this
optical correction should yield the correct \Halpha\ and $R$ band fluxes
modulo a zeropoint offset.

Figure~\ref{f:avsa} compares
the internal dust absorption in the FUV, \afuv, with that in \Halpha,
\aha, derived from the relations described above. It demonstrates 
that the internal dust corrections are broadly consistent with vectors
from observed reddening laws, especially those that account for the
differential extinction between emission lines and the stellar continuum
\citep[e.g.][]{c00}.   Since these are
transformations from the UV color and $M_R'$ respectively, these
quantities are shown on the alternate $y$ and $x$ axes.  For our adopted
dust absorption law, the $x$ axis also shows $2 A_R$ where $A_R$ is the
$R$ band absorption.  The Pearson correlation coefficient \rxy\ between
\afuv\ and \aha\ is a 0.59.  The probability to achieve this at
random from uncorrelated quantities is $4\times 10^{-12}$.  
The
average ratio, $\langle \afuv/\aha \rangle = 1.68$, is plotted as a
thick solid line in Fig.~\ref{f:avsa}.  The dashed lines show that this
is close to what is expected for the \citet{c00} starburst dust
absorption model, which includes differential absorption,
applied to population synthesis models (as described below).  This model
results in $\afuv/\aha = 1.54$ and $\afuv/(2A_R) = 1.50$.  The observed
average ratio is somewhat shallow for \Halpha\ compared to ``normal''
dust reddening laws, where dust and gas are treated the same.  For
example, using the same modeling we calculate $\afuv/\aha = 2.50$ for
the \citet{ccm89} reddening law (shown as the dash - triple dot line in
Fig.~\ref{f:avsa}).  A better agreement is found with the continuum,
$\afuv/(2A_R) = 1.60$ for \citet{ccm89} (dash - dot line).

The amplitude of the dust corrections are also consistent with what is
known about normal star forming galaxies.  Our sample has a range of
\aha\ from 0.13 to 1.8 mag with an average and median of 0.71 and 0.56
mag respectively.  We compare this to the Spitzer Infrared Nearby Galaxy
Survey (SINGS) for which galaxy averaged \aha\ can be estimated from a
weighted average of \Halpha\ and Spitzer 24$\mu$m fluxes, following the
calibration presented by \citet{c07}.  \citet{k08} do this
and find the SINGS sample has \aha\ ranging from 0 to 2.5 mag with the
median at $\sim 0.7$ mag.  That is, broadly consistent with what we
find.  For the FUV, we find \afuv\ ranging from 0 to 3.03 mag, with an
an average and median of 1.2 mag.  From Fig.~8 of \citet{gdp07} 
it can be seen that \afuv\ typically lies
between 0.6 and 2 mag in their angular diameter selected sample.
Despite the general consistency of our dust corrections with
expectations, one must always be careful of the biases dust may cause,
especially in cases like ours where the estimate of the correction is
indirect.  For example, Fig.~\ref{f:avsa} shows that relative to the
$\langle \afuv/\aha \rangle$ line, galaxies with low \aha\ ($\lesssim
0.6$ mag) tend to have high \afuv, while for high \aha\ ($\gtrsim 1$
mag) \afuv\ tends to be low compared to \aha.  In Sec.~\ref{s:dust} we
show that this difference does not have a significant effect on our
results.

\subsection{Quantities and units}

The primary quantity we work with here is the ratio of the \Halpha\ line
flux \Fha, to the FUV flux density \ffuv\ (incident flux per unit
wavelength), \Fhafuv, which has units of \AA.  We compare this to the
effective surface brightness in \Halpha\ and the $R$ band, \SHa\ and
\SR, respectively.  We show the \Halpha\ surface brightness \SHa\ in
units of W Kpc$^{-2}$ (= $5.1\times 10^{47}\, {\rm erg\, cm^{-2}\,
  s^{-1}\, arcsec^{-2}}$) and the $R$ band surface brightness in units
of $L_{R,\odot}$ Kpc$^{-2}$, where the sun's $R$ luminosity is
$L_{R,\odot} = 4.39\times 10^{22}$ W \AA$^{-1}$ ($M_{R,\odot} = 4.61$
ABmag).

\section{Stellar population models}\label{s:popmod}

\subsection{Sensitivity to IMF parameters}

\begin{figure*}
\epsscale{1.0}
\plotone{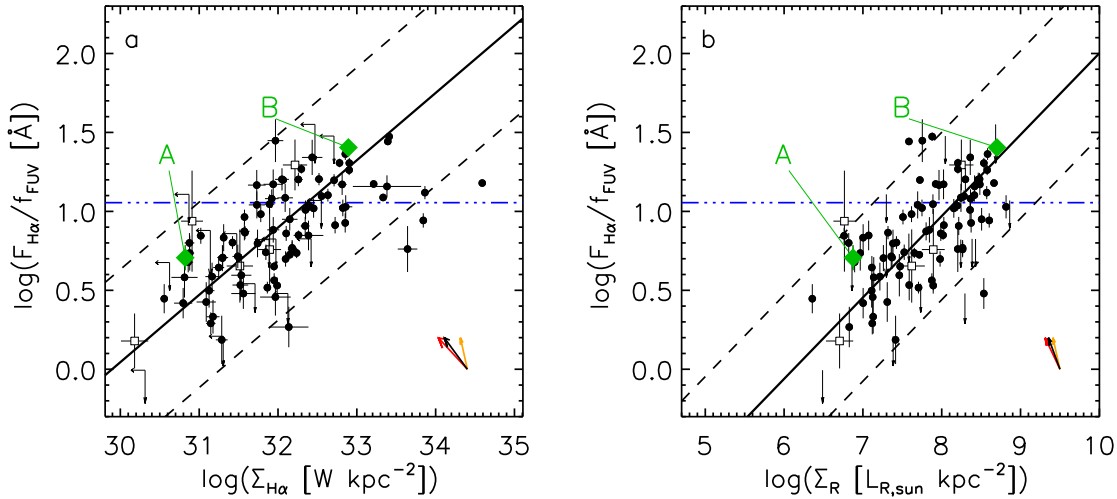}
\caption[]{The ratio of \Halpha\ line flux to FUV flux-density as a
  function of (a) the face-on \Halpha\ effective surface
  brightness, and (b)
  (right) the face-on $R$ band effective surface brightness \SR.  All
  quantities have been corrected for Galactic and internal dust
  absorption as described in the text.  The solid line shows the
  iteratively clipped least squares fit to the data, while the dashed
  lines show the final clipping limits.  The dot-dash line shows the
  \Fhafuv\ expected for our fiducial stellar population
  (eq.~\ref{e:grat}). The vectors in the lower right
  portion of the panels shows the effect of internal dust absorption according 
  to the \citet{c00} starburst attenuation law (red
  vector) the \citet{ccm89} Milky Way dust extinction law (orange
  vector), as well as our method
  (black vector).  The length of the vectors in $\log(\Fhafuv)$ in all
  cases is set to the average correction we deduce.
  Galaxies corresponding to measurements A and B are shown in
  Fig.~\ref{f:examp}. 
\label{f:hafuvsb}}
\end{figure*}

The luminosity in \Halpha\ and FUV of a stellar population arise from
different but overlapping mass ranges of stars; this drives the
sensitivity of \Fhafuv\ to the properties of the upper end of the IMF.
This is illustrated in Fig.~\ref{f:mupmods} which shows the effect
of varying \Mupp\ on the expected \Halpha\ and FUV luminosities for
stellar populations having a single slope IMF with $\Mlow = 0.1$ \Msun,
and a constant fixed SFR for the stars having $\Mstar \leq
1\, \Msun$.  The stellar population synthesis models were calculated
using the online Starburst99 package \citep[v5.1,][]{starburst99,vl05}.
They use solar metallicity Padova group evolutionary tracks
\citep{bfbc93,fbbc94a,fbbc94b,gbbc00} and model atmospheres from
\citet{phl01} and \citet{hm98}; these are the default options in Starburst99.  
Case B recombination was assumed to convert the
output of ionizing photons to an \Halpha\ luminosity \citep{o89}.  The
FUV luminosity was determined by applying the GALEX FUV transmission
curve to the spectra generated by Starburst99.  We adopt a fixed
duration $\Delta t = 1$ Gyr for the star formation so the births of O
and B stars are in equilibrium with the deaths.  Figure~\ref{f:mupmods}
shows that only stars with $\Mstar \gtrsim 3 \Msun$ contribute
significantly to the \Halpha\ or FUV luminosities and these have main
sequence lifetimes $\leq 325$ Myr
\citep{smms93,scmms93,fbbc94a,fbbc94b}. Three model sequences with IMF
slope $\gamma = -1.3, -2.35, -3.3$ are illustrated.  The central value
corresponds to the \citet{s55} slope which is consistent with more
recent determinations of Kroupa (2001, 2002). \nocite{kroupa01,kroupa02}
while the other two values are meant to represent shallow and steep
extremes to the slope.  The $y$ axis plots the luminosities relative to
the case were $\Mupp = 100\, \Msun$.  This corresponds to the highest
stellar mass determined dynamically using binary star orbits
\citep{bsuwzkssps04,rdncgsvvw04,sccms08} and is a typical \Mupp\ adopted in
the literature.  The horizontal lines indicate the masses below which
the contribution to the luminosities is 5\%, 50\%\ and 95\%\ of the
total luminosity if the true \Mupp\ is 100 \Msun.  We tabulate the
\Mstar\ values corresponding to these contributions in
Table~\ref{t:mupmods}.  

The \Halpha\ curves are very steep and closely spaced together.  This
indicates that the mass range contributing to the \Halpha\ luminosity is
more sensitive to \Mupp\ than $\gamma$.  The opposite is true for the
FUV luminosity - the curves are relatively shallow for \Mstar\ near
\Mupp\ and the shapes are very different as a function of $\gamma$ this
indicates that the mass range contributing to the FUV luminosity is more
sensitive to $\gamma$ than \Mupp.

\subsection{Fiducial stellar population model}\label{s:fmod}

From the models described above, we adopt the model with $\gamma =
-2.35$ and $\Mupp = 100$ \Msun\ as our fiducial stellar population
model.   These parameters, or similar values are
often used to characterize normal star forming stellar populations. For
example \citet{k98b}  adopted these
parameters for his SFR calibrations, although they are based on different stellar
population models \citep{mpd98}. From our fiducial model we derive the
following SFR calibrations, and compare them to the \citet{k98b}
calibrations (in parenthesis):
\begin{equation}
\frac{\rm SFR(\Halpha)}{\rm 1 \Msun\, yr^{-1}} = \frac{\LHa}{1.04\,(1.27)\times
10^{34}\, {\rm W}}, \label{e:sfrha}
\end{equation}
\begin{equation}
\frac{\rm SFR(FUV)}{\rm 1 \Msun\, yr^{-1}} = \frac{l_{\rm FUV}}{9.12\,(9.09)\times
10^{32}\, {\rm W\, \AA^{-1}}}. \label{e:sfruv}
\end{equation} Here the UV calibration of \citet{k98b} at $\lambda
= 2150$\AA\ was transformed to the pivot wavelength $\lambda_P =
1535$\AA\ of the GALEX FUV filter assuming the an intrinsic power law
FUV spectrum $f_\lambda \propto \lambda^\beta$ with UV spectral slope
$\beta = -2$.  To compare results to those that adopt the
\citet{kroupa01} IMF \citep[e.g.\ ][]{bcwtkhb04,kauffmann03b}, the SFR
estimates here should be divided by 1.5 \citep{bcwtkhb04}.  

\begin{figure*}
\epsscale{1.0}
\centerline{\hbox{\includegraphics[width=6.75cm]{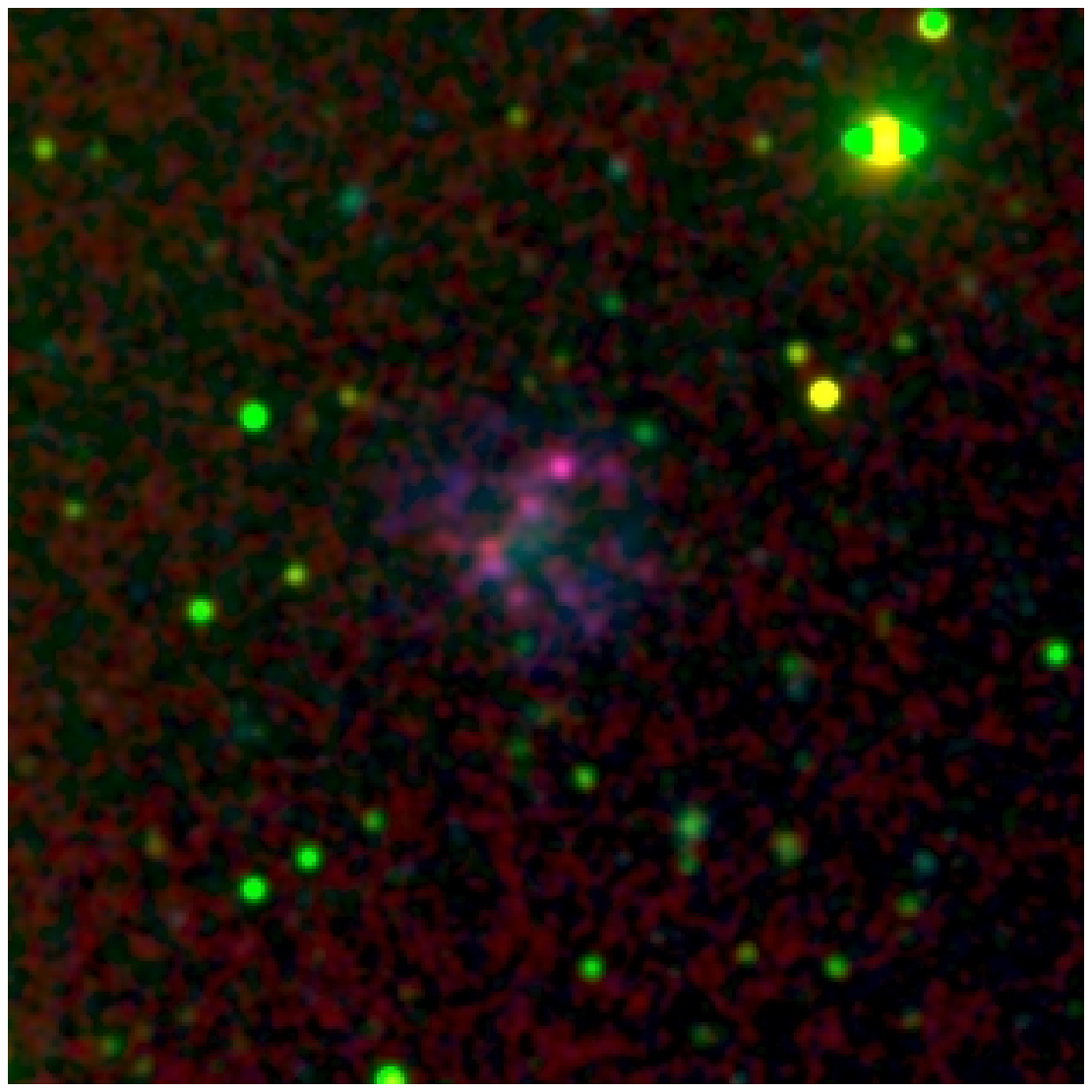}
                  \includegraphics[width=6.75cm]{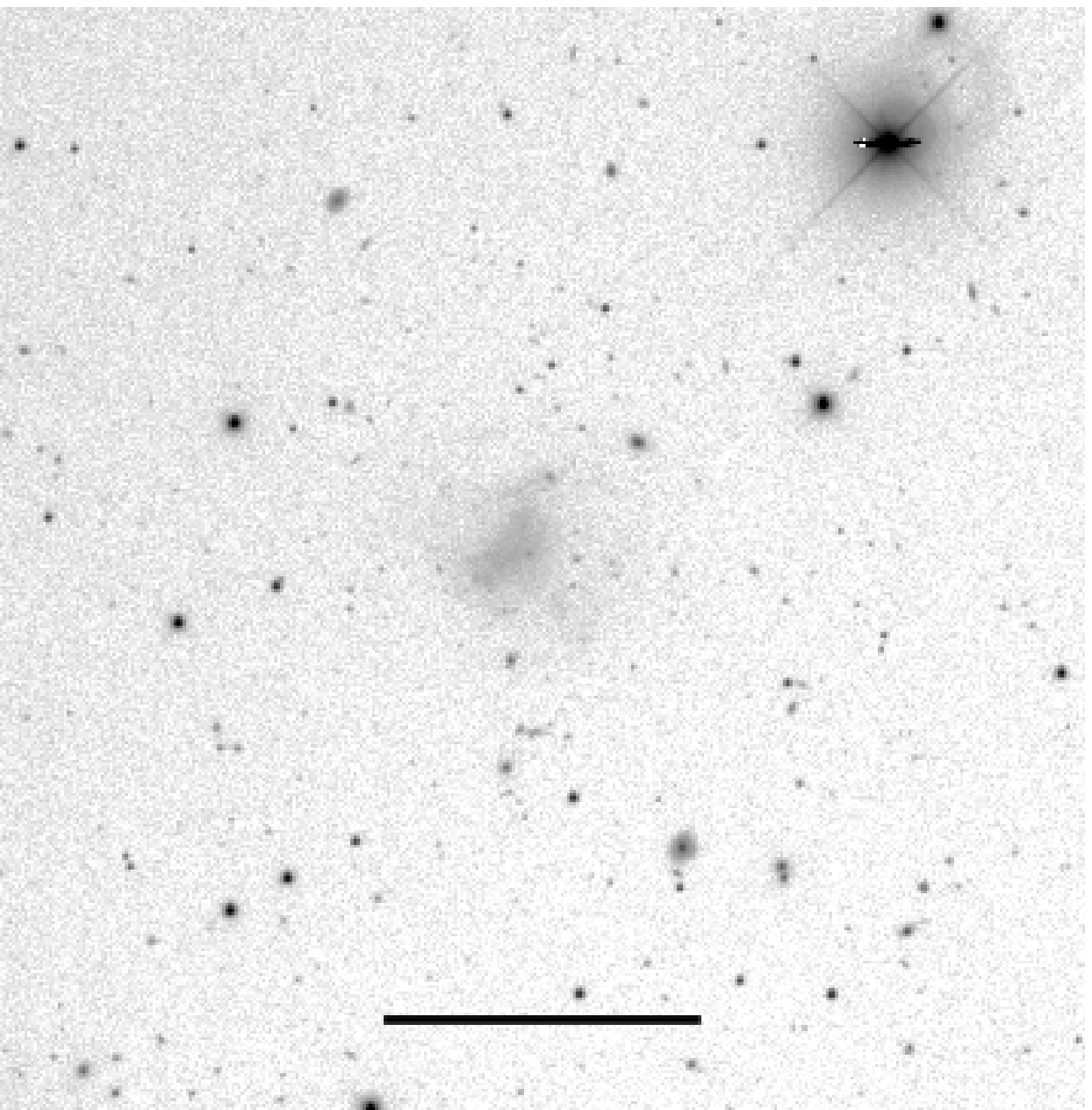}}}
\caption[]{Example galaxy A - UGCA44 (\HIPASS\ J0249$-$02) shown in
  \Halpha, $R$ band, and FUV in red, green, and blue respectively in the
  left panel, while the right panel shows the $R$ band only in an
  inverted gray scale.  The three images that comprise the right panel have
  been convolved to have the  GALEX resolution of 6.5\as, while the image
  on the left panel is shown at the original resolution of the SINGG
  data (typically $\sim 1.6''$).  The scalebar in the right panel has a
  length of 10 Kpc. \label{f:examp}}
\end{figure*}

\begin{deluxetable}{lccl}
  \tablecolumns{4}
  \tablewidth{0pt}
  \tablecaption{Properties of example galaxies\label{t:examp}}
  \tablehead{\colhead{Parameter} & \colhead{Example A} &
  \colhead{Example B} & \colhead{units}}
\startdata
Name                              &      UGCA044      &       NGC1566     &   \nl
HIPASS+                           &      J0249$-$02   &       J0419$-$54  &   \nl
Morphology                        & IB(s)m:           & (R$'_1$)SAB(rs)bc &   \nl
$\log({\cal M_{\rm HI}})$         &      8.85         &      10.19        &   \Msun\nl
$\log(L_R)$                       &   $  8.29\pm0.04$ &    $ 11.09\pm0.01$&   $L_{R,\odot}$\nl
$M_{\rm FUV}$                     &   $-14.76\pm0.04$ &    $-20.66\pm0.00$&   ABmag\nl
$r_{50}$                          &   $  2.37\pm0.05$ &    $  6.76\pm0.09$&   Kpc\nl
$\log(\Sigma_{\rm H\alpha})$      &   $ 30.82\pm0.07$ &    $ 32.89\pm0.05$&   ${\rm W\, Kpc^{-2}}$\nl
$\log(\Sigma_{\rm R})$            &   $  6.87\pm0.01$ &    $  8.70\pm0.01$&   $L_{R,\odot}\, {\rm Kpc^{-2}}$\nl
$\log(F_{\rm H\alpha}/f_{\rm FUV})$ & $  0.71\pm0.08$ &    $  1.40\pm0.05$&   \AA\nl
\enddata             
\end{deluxetable}

The ratio of eq.~\ref{e:sfruv} and eq.~\ref{e:sfrha} gives
\Fhafuv\ for the fiducial model:
\begin{equation}
\fFhafuv = 11.3\label{e:grat}
\end{equation}
We show this ratio in various plots as an indication of
the a priori expected \Fhafuv.

In order to test the sensitivity of our results to the choice of stellar
population models we also calculated models using the codes PEGASE
\citep[v2;][]{fr97,fr99} and GALAXEV \citep{bc03}.  In both cases we
adopted an IMF, metallicity, and SFH identical to the fiducial
stellar population.  Both the PEGASE and GALAXEV models use the same
Padova group sources for their evolutionary tracks as does our
Starburst99 model.  The differences are in the Stellar atmospheres.  The
PEGASE model uses a library of observed stars \citep{fr97} while GALAXEV
uses various BaSeL theoretical atmospheres as described by
\citet{bc03}.  We used our own software to 
calculate \ffuv\ as described
above, while the models provide either \LHa\ (PEGASE)
or ionizing photon flux from which we calculate \LHa\ (GALAXEV).  We
calculate an equilibrium $\Fhafuv = 10.8$, 11.8 for the PEGASE and
GALAXEV models respectively.  Hence, differences in stellar population
models can affect the \Fhafuv\ calculations at the $\sim$10\%\ level.
Metallicity also effects the model results; metallicity sensitivity is
considered in Sec.~\ref{s:metals}.

The SFR conversion factors and equilibrium \Fhafuv\ values all assume that the IMF is universal.  However, as
shown here, it is likely that the IMF is variable and often is not
consistent with our fiducial model. \citet{pwk07} examine the
implications to \Halpha\ based SFR estimates from an IGIMF that depends
on the SFR, as one would expect for star formation dominated by clusters
of finite mass.  Then the conversion factors can vary from the above by
orders of magnitude.  This scenario is discussed further in
Sec.~\ref{s:stoc} and Sec~\ref{s:implic}.

\section{Results}\label{s:results}

Figure~\ref{f:hafuvsb} shows our primary result: a strong correlation
between \Fhafuv\ and surface brightness in both \Halpha\ (panel a), and
in the $R$ band, \SR\ (panel b).  One striking aspect of the Figure is
that more than half the sample has \Fhafuv\ below normal.  Except for 
possibly one weak \Halpha\ detection,
and one galaxy with an upper limit to \Fhafuv, this includes
all of the LSB galaxies having $\log(\SHa) \leq 31.7$ or
$\log(\SR) \leq 7.5$.  
Ordinary least squares bisector fits were performed to the data, with
iterative clipping of the points having residuals outside of $\pm 2.5
\sigma_y$ from the best fit line yielding
\begin{equation}
\log(\fFhafuv) = (-12.75 \pm 1.02) + (0.43\pm 0.03)\log(\SHa),\label{e:corrsha}
\end{equation}
\begin{equation}
\log(\fFhafuv) = (-3.16 \pm 0.28) + (0.52\pm 0.03)\log(\SR).\label{e:corrsr}
\end{equation}
These are shown as solid lines in Fig.~\ref{f:hafuvsb} while the
clipping limits are shown as dashed lines.  Data with only \Halpha\
upper limits were not included in any of our fits. The quality of the
correlations and fits are very similar with each having a Pearson's
correlation coefficient $\rxy = 0.65$, a probability of $\sim 4\times
10^{-12}$ that the correlation could have occurred from random data, and
an rms dispersion of 0.22 dex about the $\log(\Fhafuv)$ residuals (after
clipping).

Optical surface brightnesses are not the only quantities that correlate
with \Fhafuv.  This is illustrated in two ways.  First, in
Fig.~\ref{f:examp} we show example images of galaxies at either end of
the correlations as indicated in Fig.~\ref{f:hafuvsb}; some properties
of these two examples are given in Table~\ref{t:examp}.  This
illustrates the correlation with morphology along this sequence.
The LSB example galaxy, UGCA44 is a dwarf irregular galaxy without
coherent structure.  The High Surface Brightness (HSB) example,
NGC~1566, the brightest member of the Dorado group, is a strong regular
two arm spiral galaxy.  These two examples are typical of the
morphological variation along the sequence, and consistent with what is
known about the sequence of star forming galaxies: LSB sources are
typically low luminosity irregular systems, HSB galaxies tend to be high
luminosity spirals (as well as some blue compact dwarfs), galaxies with moderate surface brightness are a
mixture of late type spirals and brighter irregulars.  

\addtocounter{figure}{-1}
\begin{figure*}
\epsscale{1.0}
\centerline{\hbox{\includegraphics[width=6.75cm]{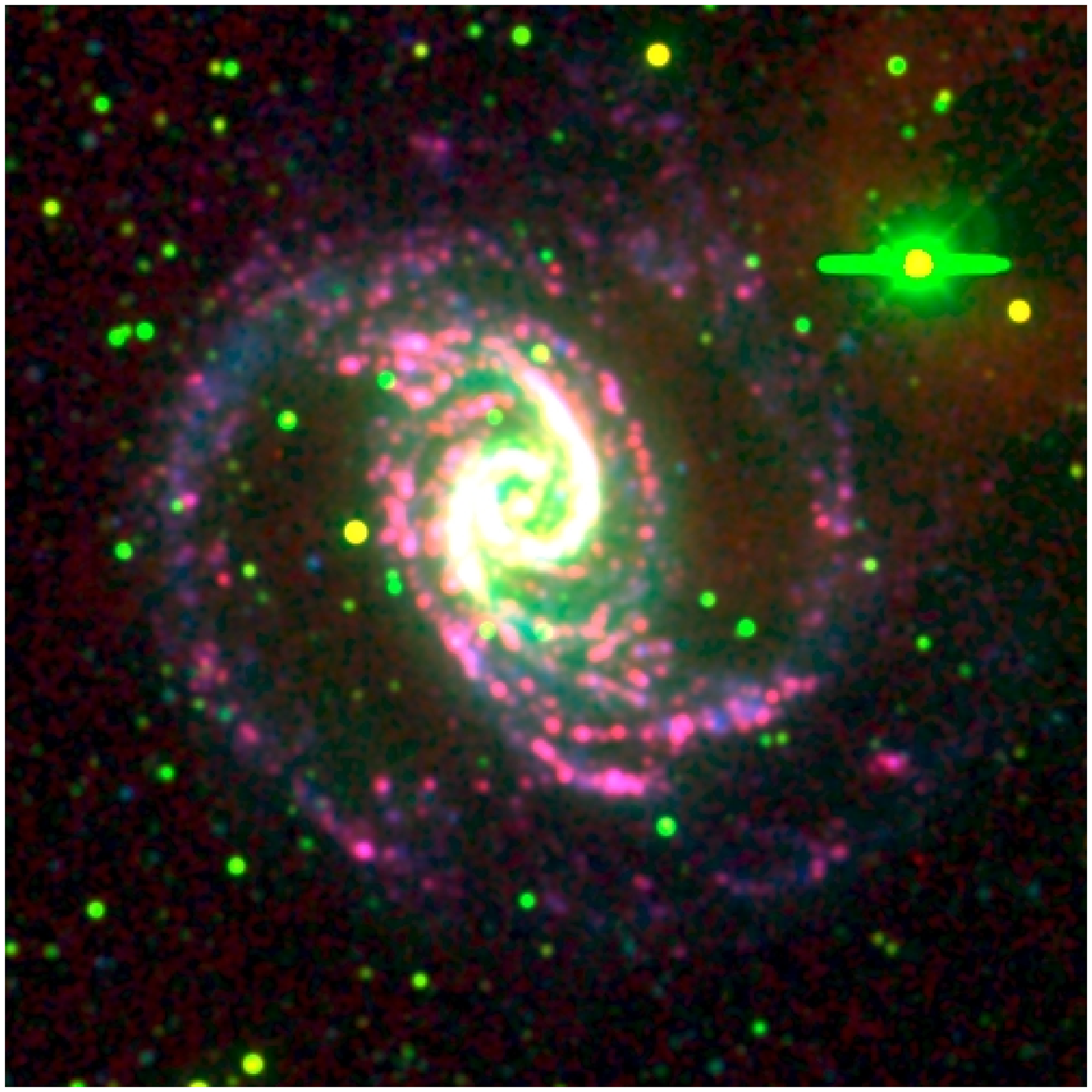}
                  \includegraphics[width=6.75cm]{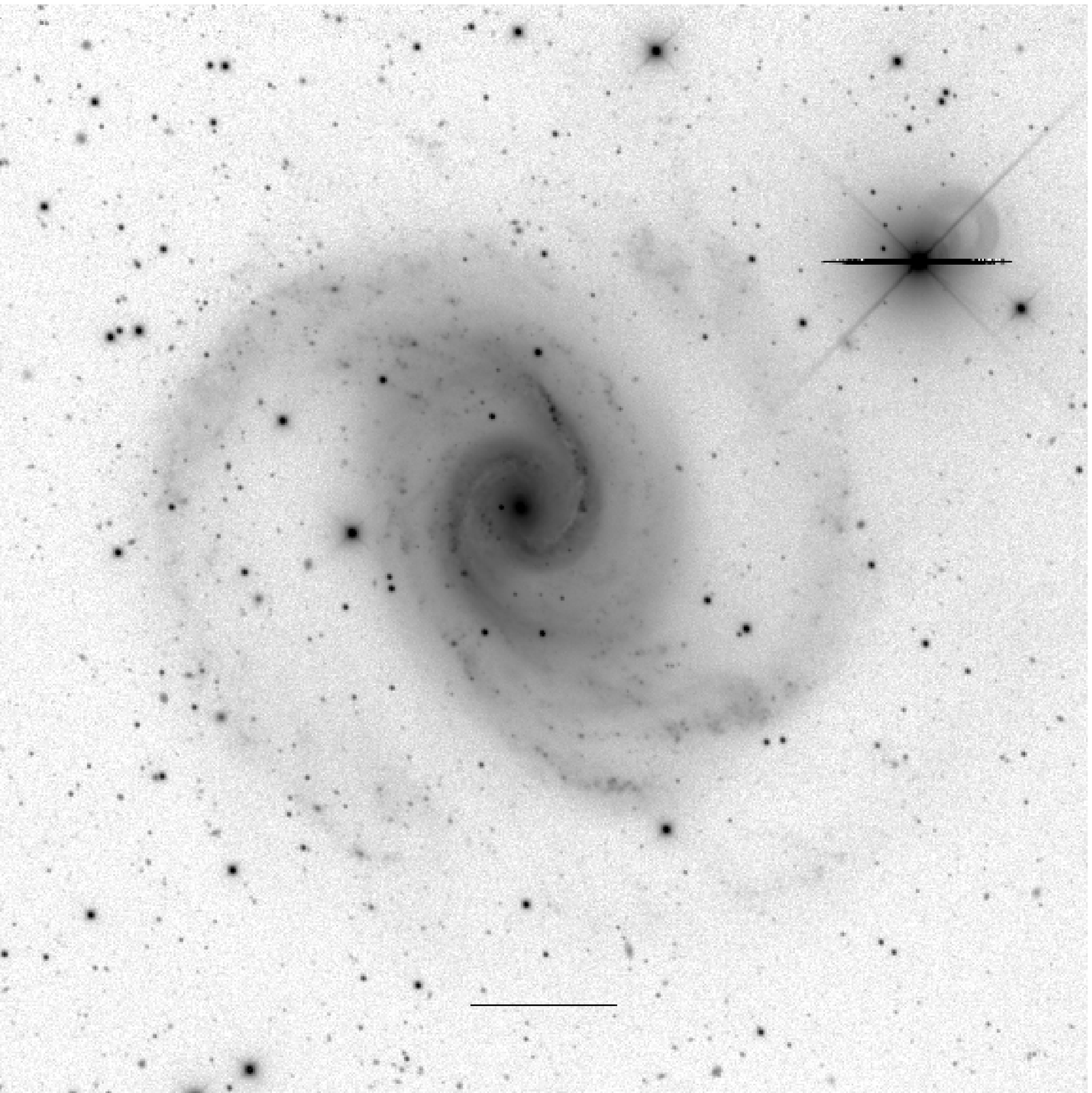}}}
\caption[]{continued. Example galaxy B - NGC1566 (\HIPASS\ J0419$-$54).
  The right and left panels have the same display levels
  as the other example. The projected area of these images is larger,
  while the scalebar length remains 10 Kpc. }
\end{figure*}

Second, \Fhafuv\ correlates with other global quantities such as total
luminosity \LR\ and rotational velocity \Vrot\ (determined from the \HI\
line widths as outlined in H06), which is shown in Fig.~\ref{f:hafuvlm}.
Only \HIPASS\ sources corresponding to single ELGs with optical axis
ratio $a/b > 1.4$ are plotted in panel b, in order to limit the plot to
galaxies likely to be dominated by rotation. Using the same fitting
algorithm as described above, we find
\begin{equation}
\log(\fFhafuv) = (-1.64 \pm 0.15) + (0.27\pm 0.01)\log(\LR),\label{e:corrlr}
\end{equation}
\begin{equation}
\log(\fFhafuv) = (-0.97 \pm 0.17) + (0.96\pm 0.09)\log(\Vrot).
\end{equation}
These correlations are weaker than those shown in
Fig.~\ref{f:hafuvsb}, having correlation coefficients of 0.61 (0.57)
while the (clipped) rms scatter in the $\log(\Fhafuv)$ residuals is 0.19
(0.32) dex in panel a (b).  Nevertheless, the probability that these
correlations could have come from randomly drawn data is still low:
$10^{-10}$ for the correlation with log(\LR), and $3\times 10^{-4}$ for the
correlation with log(\Vrot) (low, but much higher than the other correlations
shown here because of the smaller sample of elongated single ELGs).

The range of luminosity and mass in these correlations are too large for
galaxies to evolve from one end of the correlation to the other.
Instead, \Fhafuv\ is one of the global properties (although not a
fundamental one) that vary
regularly along the sequence of late type galaxies.  Other well known
correlations along this sequence include the \citet{h26} morphological
sequence, the Tully-Fisher (1977)\nocite{tf77} relation, the
mass-metallicity relation \citep{tremonti04} and the Universal Rotation
Curve \citep{ps91,pss96}. 
The main physical parameter that drives most
of these correlation appears to be halo-mass.  However, the correlations
in Fig.~\ref{f:hafuvsb} are stronger than those in Fig.~\ref{f:hafuvlm}
indicating that star formation intensity and stellar mass density are
more important for driving the \Fhafuv\ correlations than halo properties.

Both the range of \Fhafuv\ values, covering over a factor of ten, and
the correlation with surface brightness have major implications, as
discussed in Sec.~\ref{s:implic}.

\section{Parameters affecting \Fhafuv}\label{s:ratpars}

Physical properties that can affect \Fhafuv\ include dust absorption,
the detailed SFH, the porosity of the interstellar medium, stochastic
effects from the number of stars, the metallicity of the stellar
populations, and the properties of the IMF.  We now consider, in turn,
which of these could drive the observed correlations, saving the IMF for
last because it is the only option we can not rule out.

\subsection{Dust}\label{s:dust}

The dust corrections are too small and in the wrong sense to have
spuriously caused the observed correlations.  Moreover, even
without any dust corrections, a large fraction of the sample have too
few ionizations for the observed UV light compared to expectations for
normal star forming populations.  These points are illustrated in
Fig.~\ref{f:nodust} which is similar to Fig.~\ref{f:hafuvsb}, but here
the quantities are shown before internal dust correction (and denoted
with  a prime: \SHar, \SRr\ and \Fhafuvr).  We see
strong correlations between the quantities 
even before dust correction.  Here $\rxy = 0.66$ (0.69) for the
correlation between \Fhafuvr and \SHar (\SRr), while the
probability that these are spurious correlations from random data is
$7\times 10^{-13}$  ($3 \times 10^{-14}$).  
The best fits to these data are
\begin{equation}
\log\left(\fFhafuvr\right) = (-11.32\pm 1.14) + (0.39\pm 0.04)\log(\SHar),\label{e:corrshar}
\end{equation}
\begin{equation}
\log\left(\fFhafuvr\right) = (-3.36\pm 0.32) + (0.58\pm 0.04)\log(\SRr).\label{e:corrsrr}
\end{equation}
As before, the fits are OLS bisector fits with an iterative $2.5\sigma$
clipping.  The scatter about the fit is 0.22 (0.23) dex in
$\log(\Fhafuvr)$ for the \SHar\ (\SRr) data, nearly identical to the
scatter seen in Fig.~\ref{f:hafuvsb}.  

\begin{figure*}
\epsscale{1.0}
\plotone{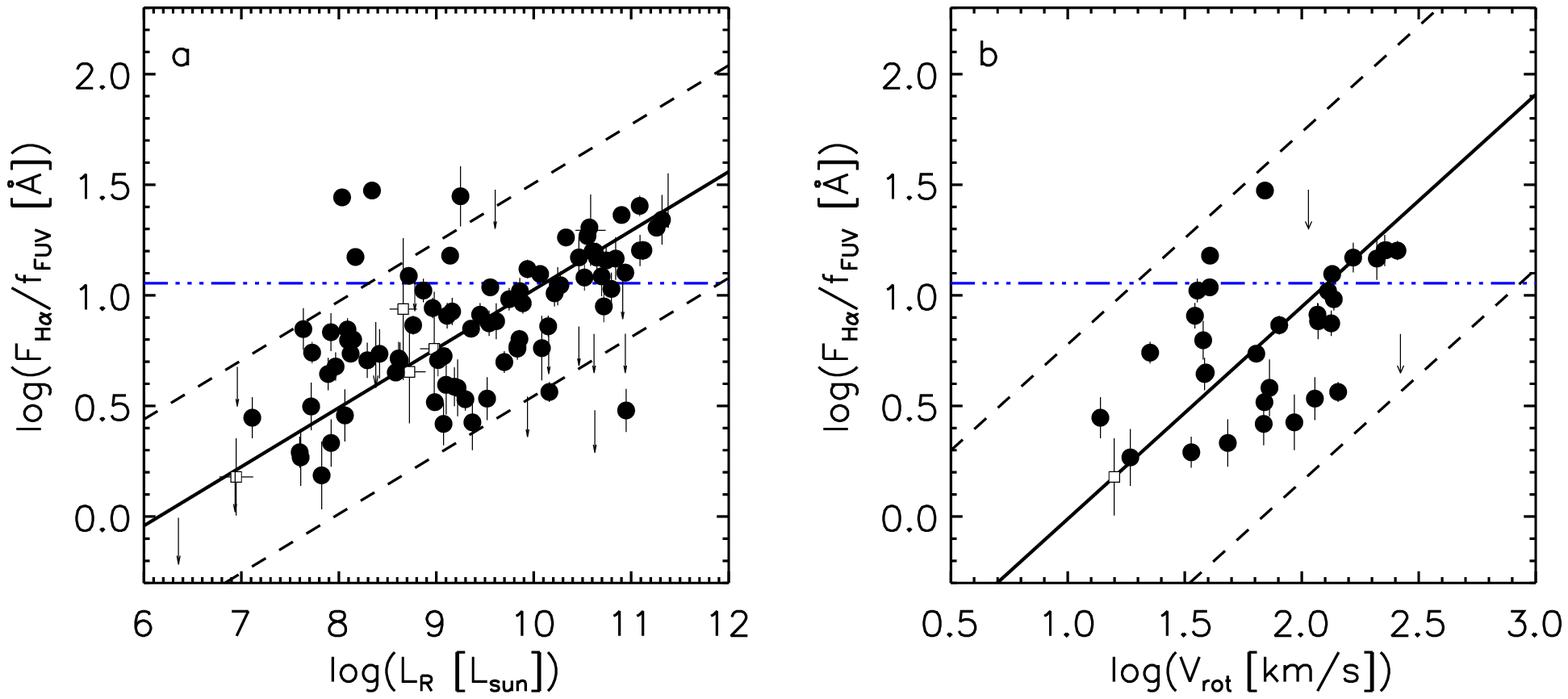}
\caption[]{\Fhafuv\ as a
  function of (a) $R$ band luminosity and (b) rotational
  velocity. The symbol and line types have the same meaning as in
  Fig.~\ref{f:hafuvsb}.  \label{f:hafuvlm}}
\end{figure*}

There are several reasons why reddening correction does not greatly
effect the results.  First, the average dust correction, shown by the black arrows in Figs.~\ref{f:hafuvsb} and
\ref{f:nodust}, is not large compared to the range of the correlations. 
Second, The difference in the dust correction between
the high and low surface brightness ends is also not that large.  This
is shown by dividing the sample into quartiles of \Fhafuvr.  The
quartile boundaries are at $\log(\Fhafuvr) = 0.91$, 1.04, and 1.26 and
shown with dotted lines in Fig.~\ref{f:nodust}.  The average of the
uncorrected points within the quartiles are shown as large pink
triangles, while the average of the same data points after dust
correction is shown with large cyan diamonds; thick lines show OLS-bisector
fits to these average points.  Systematic changes in the $\afuv/\aha$
ratio noted in \S\ref{s:cdust} are also not large enough to seriously effect the
results.

Figure~\ref{f:nodust} also shows that nearly half the sample have
\Fhafuv\ below the value for our fiducial stellar population model.  The
first quartile has $\langle\log(\Fhafuvr)\rangle = 0.73 \pm 0.10$, where
here the error is the average error on $\log(\Fhafuvr)$.  Hence, the
galaxies in this quartile have $\log(\Fhafuvr)$ values that are on
average 3$\sigma$ lower than the fiducial model even before dust
correction.

\begin{figure*}
\epsscale{1.0}
\plotone{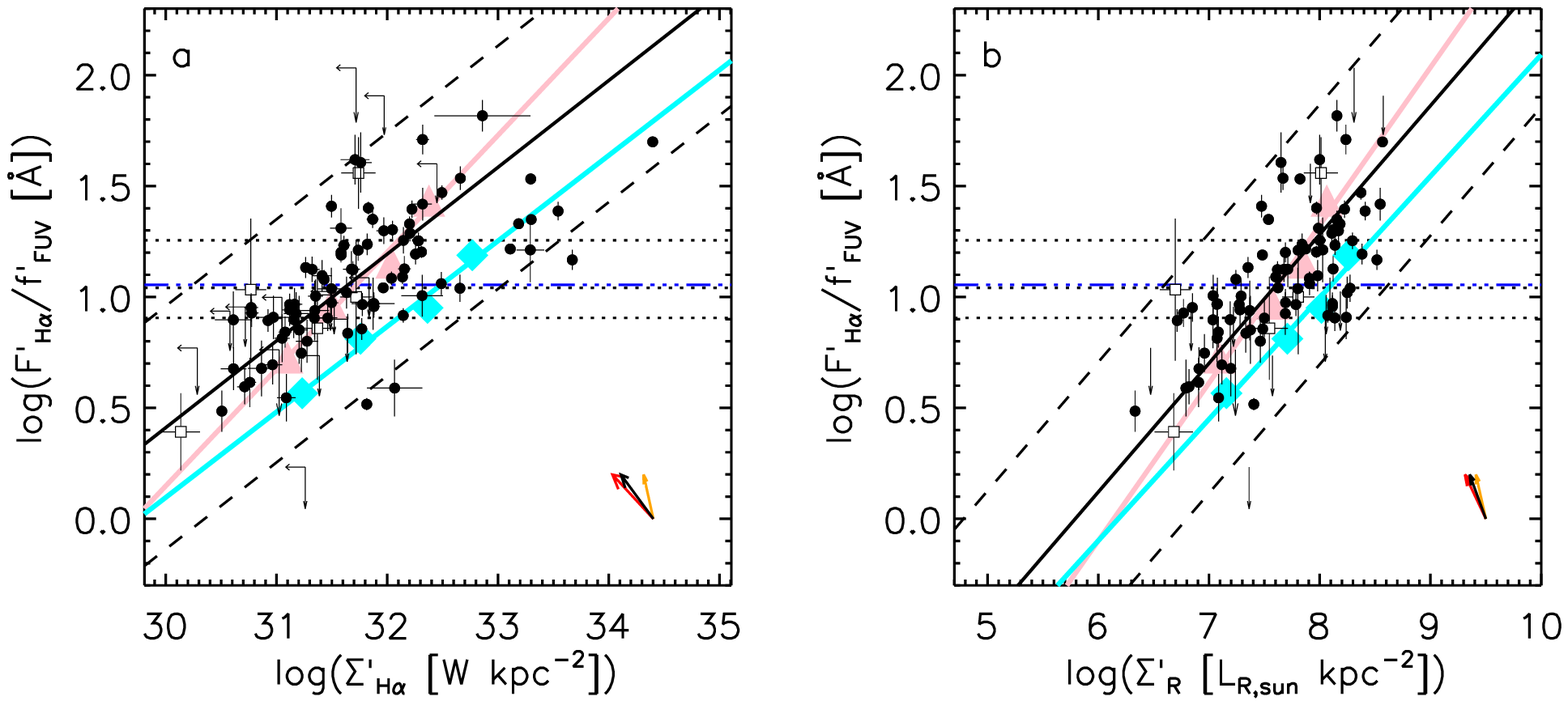}
\caption[]{The quantities shown in Fig.~\ref{f:hafuvsb} are plotted here
  prior to dust absorption correction, and hence are denoted with prime
  ($'$) symbols.  The dotted lines separate the sample into quartiles of
  $\log(\Fhafuvr)$. The pink triangles show the average qauntities of
  the data points in each quartile, while the blue diamonds shows the
  average of the same points after dust correction.  The thick pink and
  cyan lines are least squares fits to the respective average
  points. The meaning of the other 
  symbols and lines are the same as in Fig.~\ref{f:hafuvsb}.\label{f:nodust}}
\end{figure*}

Could it be that we are {\em under\/} correcting for dust absorption?
This seems unlikely, because the amount of dust needed to flatten the
relationships shown in Fig.~\ref{f:hafuvsb}, is implausibly high and
would only exacerbate the problem of low \Fhafuv\ values.  To flatten the
relationship we consider the amount of dust required to bring the
quartile $\langle\log(\Fhafuvr)\rangle$ values equal to that of the
first quartile.  In other words, this assumes that the first quartile is
virtually unaffected by dust.  Since dust almost certainly effects all
measurements this will produce an underestimate of the dust correction
needed.  In the second, third, and fourth quartiles
$\langle\log(\Fhafuvr)\rangle = 0.97$, 1.15, 1.43 respectively.  Using
$\langle\afuv/\aha\rangle = 1.68$ (Sec.~\ref{s:cdust}) then to lower these to
the value observed in the first quartile requires $\aha = 0.9$ 1.6, and
2.6 mag, respectively, and $\afuv = 1.5$, 2.6, and 4.4 mag,
respectively.  This is equivalent to an overall average $\aha = 1.3$ mag
and $\afuv = 2.1$ mag, or about twice the value estimated in Sec.~\ref{s:cdust}.

Much of the scatter in the observed correlations may come from
uncertainties in the internal dust correction.  From the scatter in the
IRX-$\beta$ fits of \citet{sunggatlas} and \citet{cbbgbgsmm06} combined
with the relationship between IRX and FUV dust absorption $A_{\rm FUV}$
of \citet{buat05} we determine that about 0.14 dex uncertainty in \ffuv\
results from the IRX-$\beta$ relation.  The correction to \Fha\ for dust
absorption is tied to Balmer Decrement measurements. \citet{kgjd02} show
that after dust correction the SFR estimated from \Halpha\ and FIR
emission agree well in the calibrating Nearby Field Galaxy Sample.
Using the online version of their table 1, we find a dispersion in their
mean ${\rm log(SFR_{\rm FIR}/SFR_{\rm H\alpha})}$ of 0.16 dex (after an
iterative 2.5$\sigma$ clipping).  Assuming this is the intrinsic scatter
due to dust correction affecting \Fha, and that it is uncorrelated with
the scatter in \ffuv, then the uncertainty in log(\Fhafuv) due to dust
may be as high as 0.21 dex.  This is very close to the observed scatter
in Figures~\ref{f:hafuvsb} and \ref{f:hafuvlm}.  However, this might be
an overestimate of the scatter due to dust since ${\rm SFR_{\rm
H\alpha}}$ estimates of Kewley et al.\ are also affected by the IMF.

Driver et al.\ (2007,2008) \nocite{dptlgad07,dptglb08} 
find strong inclination dependent differences in the
luminosity functions of the bulge and disk components of galaxies which
they attribute to dust absorption.  While inclination effects may be
seen in, and removed from, the UV with our color based \afuv\ estimate,
our \aha\ estimate is just based on luminosity and any inclination term
should remain and effect or \Fhafuv\ estimates. However, we found
insignificant correlation coefficients $\rxy = 0.13$, 0.01 for the
residuals of the fits in Fig.~\ref{f:hafuvsb} panel (a) and (b),
respectively, and the axial ratio ($a/b$) of the sources in the optical,
demonstrating that inclination induced dust absorption is not
significantly contributing to the scatter.  Further work is required to
determine what is causing the scatter in these relationships.

\subsection{Star Formation History}\label{s:sfh}

Sharp changes in the SFR can strongly effect \Fhafuv\ due to the
different lifetimes of O and B stars.  To explore the effects of a
variable SFR on our correlation we built photometric models of a
temporary increase, or ``burst'', and decrease, or ``gasp'', on an
otherwise constant SFR population buildup.  The models were created
using Starburst 99 to calculate the evolution of the spectrum of a
simple stellar population (i.e.\ instantaneous starburst). The evolution
of the relevant luminosities (\Halpha, FUV and $R$) for an arbitrary SFR
as a function of time, $t$,  were calculated from the simple stellar population
results using a separate program.  These models employed the same
fiducial IMF outlined in Sec.~\ref{s:fmod}.

\begin{figure}
\epsscale{1.1}
\plotone{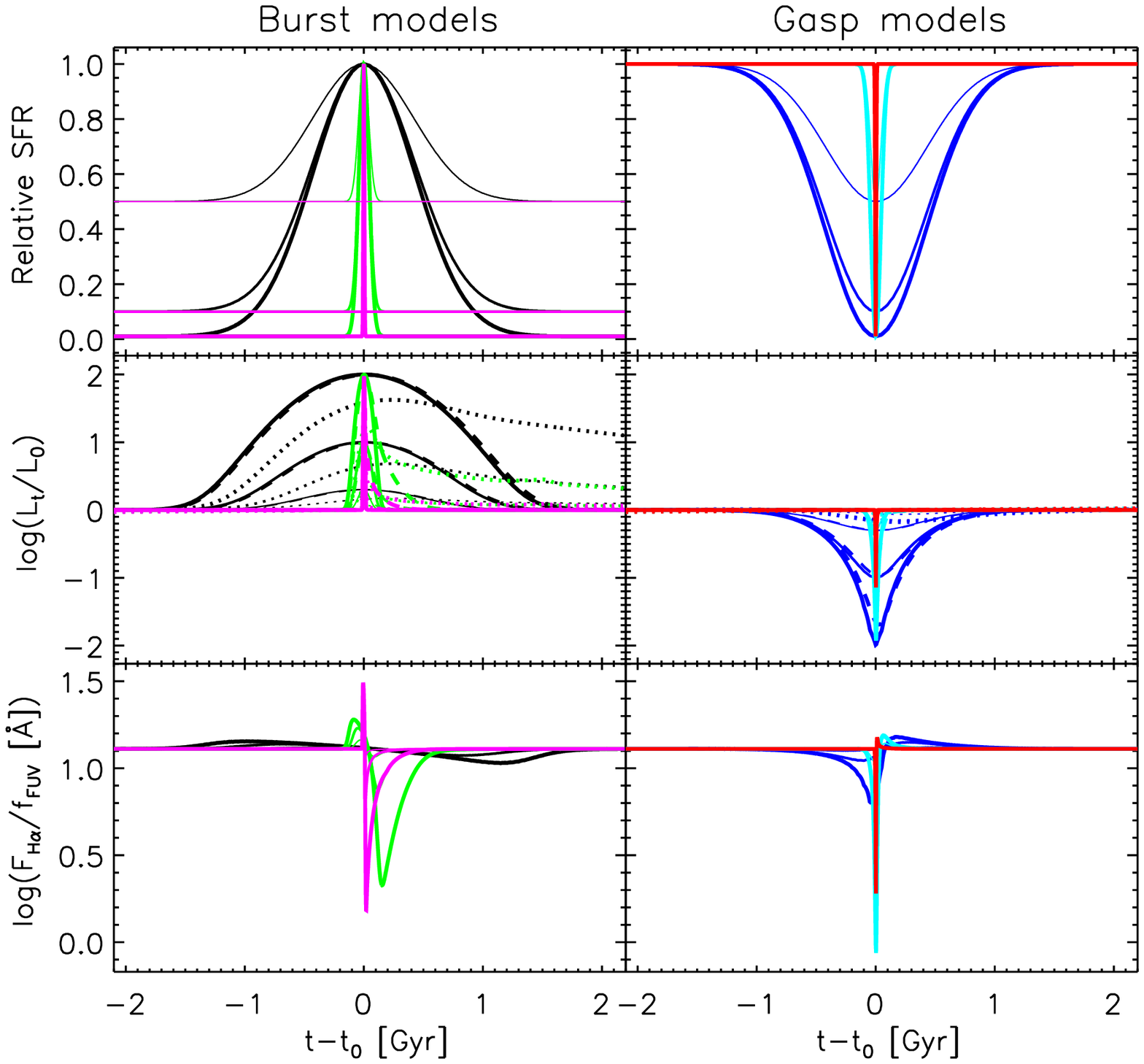}
\caption[]{Models of the photometric evolution of ``burst'' and
  ``gasp'' SFHs are shown in the left and right panel repectively.
  The top panels shows the relative star formation rate for the adopted
  Gaussian profile SFH.  The $x$ axis on all panels is the time relative
  to the event center $t_0$.  Three FWHM durations $\Delta t = 10, 100, 1000$
  Myr are displayed using different colors: magenta, green, and black
  respectively for burst models; and red, cyan, and blue, respectively
  for gasp models.  
  Three relative amplitudes, defined as SFR maximum:minimum are
  shown - 2:1, 10:1, 100:1 and distinguished with increasing line
  thickness.  The color and thickness coding are repeated in the
  remaining panels and the next figure.  The middle panel shows the
  evolution of luminosity relative to the pre-event luminosity (defined
  as $t_0 - 2\Delta t$) in the \Halpha, FUV, and $R$ band distinguished
  using solid, dashed and dotted lines respectively.  The bottom panels
  show the evolution $\log(\Fhafuv)$. 
\label{f:bg_time}}
\end{figure}

Figure~\ref{f:bg_time} shows the burst and gasp SFHs that we modeled,
the evolution of the fluxes \Fha, \ffuv, and \fr, and the evolution of
the ratio \Fhafuv.  The models are relative in the sense
that the absolute SFR is not important and is set by adding 
appropriate zeropoints.  Hence the SFR and fluxes are shown relative to
those before the burst or gasp. The functional form that we adopt is a
constant SFR with a Gaussian enhancement or depression representing a
burst and gasp, respectively.
The model parameters are the FWHM duration, $\Delta t$,
of the event, the ratio of maximum to minimum SFR, $A$, and the time
$t_0$ of the center of the event. All models have a finite
base SFR. We do not allow the SFR to completely turn-off because
our sample shows that almost all late type galaxies have at least some
recent high mass star formation (M06).   For the
burst models, the maximum SFR occurs during the event and the minimum is
the base level, while for gasp models, the minimum SFR occurs during the
event and the maximum is the base level.   All models have $t_0 = 10$
Gyr.  Absolute time is important when considering $R$ band fluxes; our
choice places the burst
or gasp at roughly the Hubble time.  

We explored a range of
model parameters and found that the salient features of these models can
be illustrated with models having $\Delta t = $ 10, 100, 1000 Myr, and
$A = 2, 10, 100$.  Figure~\ref{f:bg_time} shows that very large
excursions in \Fhafuv\ are possible, especially for large $A$ and short
$\Delta t$.  The $\Delta t = 1$ Gyr models produce only very weak
excursions in \Fhafuv\ because the timescale is long compared to the
lifetimes of both the O and B stars.

\begin{figure*}
\epsscale{1.0}
\plotone{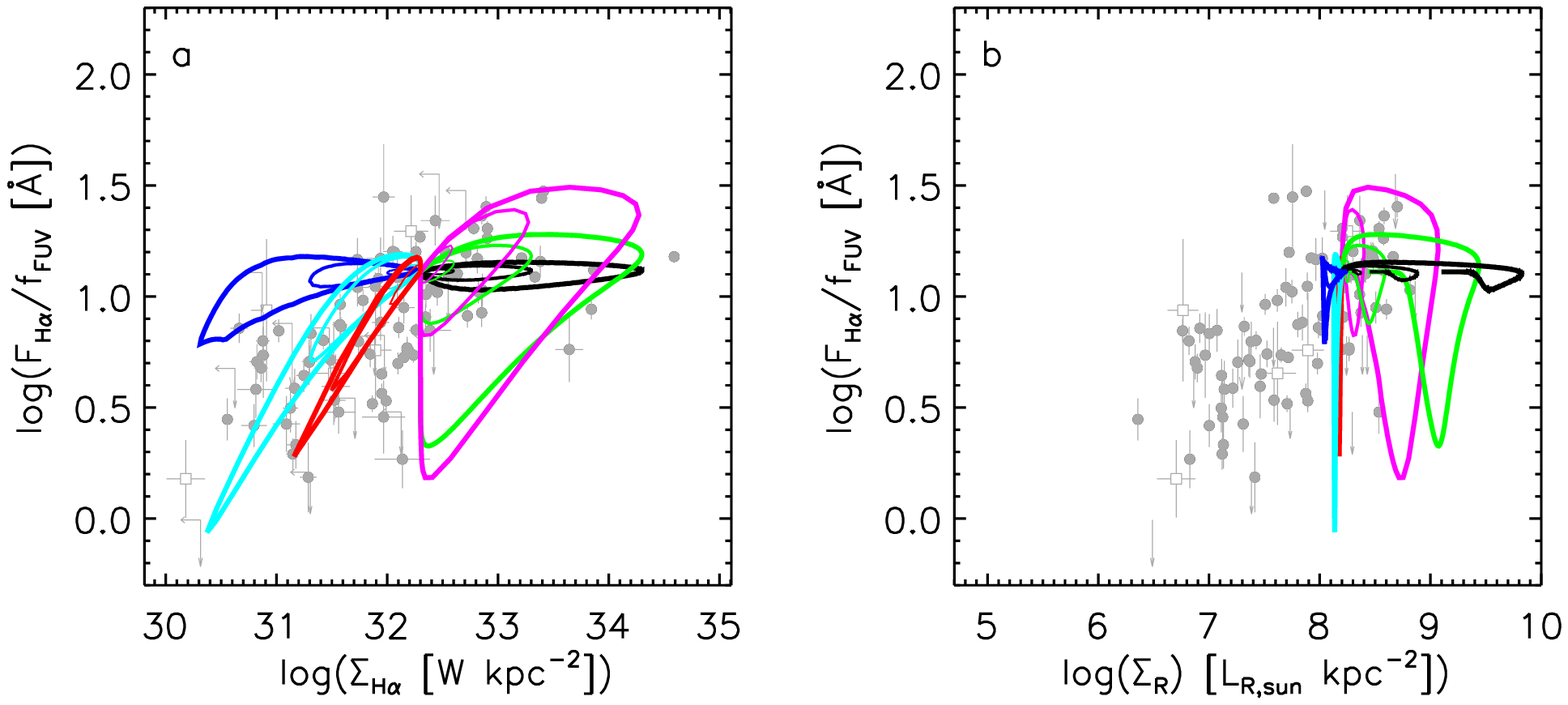}
\caption[]{Tracks of \Fhafuv\ versus \SHa\ (panel a) and \SR\ (panel b)
  made by the burst and gasp SFHs shown in Fig.~{\ref{f:bg_time}} using
  the same convention for line color to distinguish event duration and
  line thickness to represent event amplitude.  The data is
  the same as in Fig.~\ref{f:hafuvsb}, but displayed in gray so the
  model lines show up clearly.  The model trajectories converge on the
  pre-event values, for which we have adopted \SHa\ and \SR\ zeropoints to
  be consistent with typical galaxies having the equilibrium \Fhafuv. 
  }{\label{f:bg_hafuvsb}}
\end{figure*}

Figure~\ref{f:bg_hafuvsb} plots the tracks the models make in the \Fhafuv\
versus \SHa\ and \SR\ planes.  All models converge at our adopted
surface-brightness zeropoints and $\log(\Fhafuv {\rm [\AA]}) = 1.055$,
the equilibrium value (eq.~\ref{e:grat}). We chose $\log(\SHa {\rm [W\, Kpc^{-2}]}) = 32.3$ and $\log(\SR
{\rm [L_{R,\odot}\, Kpc^{-2}]}) = 8.2$ as the zeropoints corresponding
to the surface brightness immediately prior to the burst or gasp ($t
= t_0 - 2\Delta t$) since these values are near the middle of the pack
with respect to other galaxies having similar \Fhafuv.  We also ran the
models with the same IMF, metallicity and SFHs but based on PEGASE model calculations.  Other
than the zero-point offset noted in Sec.~\ref{s:fmod}, these models make
indistinguishable tracks from those shown in
Fig.~\ref{f:bg_hafuvsb}. 

Both the burst and gasp models can produce large correlated excursions
in \SHa\ and \Fhafuv\ mimicking the observed correlations.  This is
especially true for the gasp models.  However, in cases of large $A$ and
short $\Delta t$ the tracks can enter unpopulated regions
of the \Fhafuv\ versus \SHa\ diagram during the post-burst phase as 
\Fhafuv\ drops and \SHa\ fades to its
original value.  Fine tuning of the SFH to have a longer duration to the turn-off phase
may alleviate this problem.  The \Fhafuv\ versus \SR\ plane is more
difficult to explain with burst or gasp models.  This is because short
$\Delta t$ events do not last long enough to significantly change \SR,
while the long $\delta t$ events that can cause large \SR\ excursions have
little effect on \Fhafuv.  Thus a SFH including a
recent burst or gasp can not simultaneously account for the correlations
of \Fhafuv\ with both \SR\ and \SHa.  If bursts and gasps cause the
spread of \Fhafuv\ then the SFH of galaxies must be synchronized so that
LSB galaxies would all be in a gasp or post-burst phase.  We can
think of no physical mechanism that can naturally result in such a
contrived scenario.

\begin{figure}
\epsscale{1.1}
\plotone{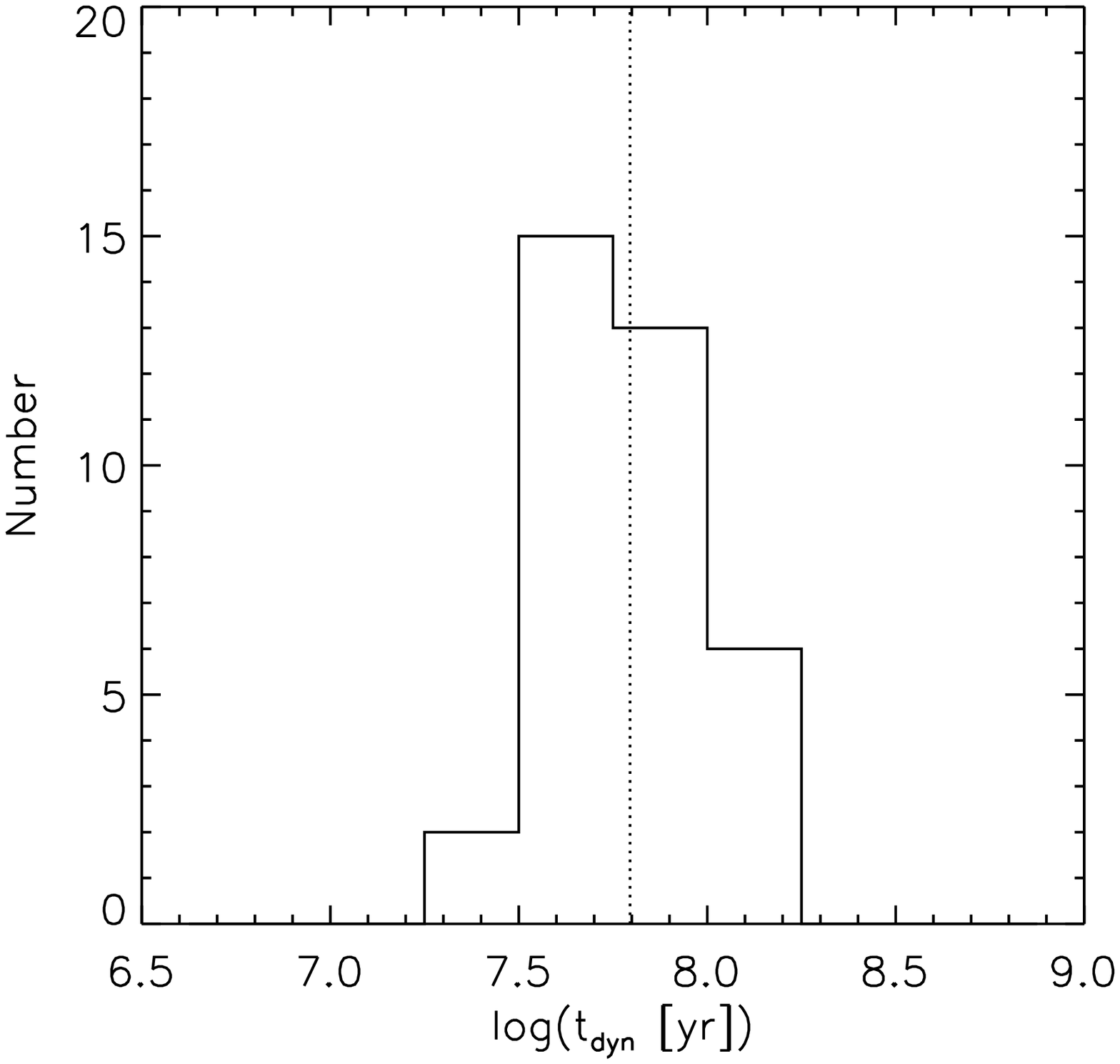}
\caption[]{Distribution of dynamical times of the sample galaxies
  consisting of single ELGs having $a/b > 1.4$. The dotted line shows
  the median of the distribution. }{\label{f:tdyn}}
\end{figure}

While the burst and gasp models can produce large excursions in \Fhafuv,
this requires short $\Delta t$ and large $A$ events.  What amplitudes
and durations are likely for intermittent star formation?  Causality
dictates that the event duration should not be shorter than the
dynamical time, \tdyn\ since it is the timescale for disturbances (such
as the build-up or removal of ISM) to travel through a galaxy.
Figure~\ref{f:tdyn} displays the distribution of \tdyn\ calculated from
the \HI\ profile width for the galaxies in our sample with axial ratio
$a/b > 1.4$.  These calculations follow the method of H06 for calculating
\Vrot\ \HI\ line width and adopt $\tdyn = \torb/\sqrt{3\pi}$, where
$\torb = 2\pi r_{50}/\Vrot$.  This definition effectively adopts a
flat rotation curve which is evaluated at the \Halpha\ $r_{50}$, 
and represents an average \tdyn\ for high mass star formation in
the galaxies.  For rotation curves that are rising at $r_{50}$, this
definition underestimates \tdyn.  From Fig~\ref{f:tdyn} it can be seen
that the median $\tdyn \approx 60$ Myr.  Numerical simulations of
starbursts indicate durations as low as $\sim 60$ Myr for the final
sharp peak in the SFR during a merger, with the peak amplitudes on the
order of 10-80 times the base level \citet{mh94a,mh94b,mh96}.  These
simulations show that enhanced star formation during a merger lasts
several 100 Myr.  Mergers are dramatic rare events and our sample shows
very few examples of these ($< 10$\%, M06).  Normal variations in SFR
should be weaker and longer.  For example, morphological and SED analyses
of starbursting dwarf galaxies suggest $A \lesssim 3$ and $\Delta t$ in
the range of several 100 Myr \citep{mfdc92,p96a,p96b,mmh99}.  Our models
indicate that such events may add scatter to our correlations but can
not explain the large range in \Fhafuv.

\subsection{Porosity of the Interstellar medium}\label{s:fesc}

The observed \Fhafuv\ correlations could also occur if the escape
fraction of ionizing photons, \fesc, inversely correlates with surface
brightness.  This explanation would require that the galaxies in the
lowest quartile of \Fhafuv\ are losing $\gtrsim 50$\%\ of their ionizing
flux compared to our fiducial model.
Escaping ionizing photons have been detected in UV bright star forming
galaxies at a redshift $z \sim 3$ \citep{spa01,sspae06,iimfhkaybd08}. In
the local universe, direct measurements of \fesc\ have been limited to
HSB starburst galaxies and have mostly been upper limits in the range
$\fesc < 0.05 - 0.1$ \citep{lfhl96,hjd97,dblmksg01}.  The strongest
claim for a direct detection of $\fesc \sim 0.04 - 0.11$ in Haro 11 by
\citet{bzaamo06} is disputed by \citet{ghsdsohap07} who find $\fesc
\lesssim 0.02$ for the same data.  The best case
for escaping ionizing photons from a normal galaxy comes from the very
faint \Halpha\ emission detected in the Magellanic Stream and High
Velocity Clouds which are thought to be partially ionized by the disk of
our Galaxy \citep{bm99,pbvgfm03}.  These measurements require about 6\%\
of ionizing photons to escape normal to the Galactic plane, or, about
$\fesc \approx 1\%\ - 2\%$ when isotropitized over all angles.  No direct
measurements of \fesc\ from LSB galaxies have been attempted.  Hence we
can not directly rule out the possibility that \fesc\ inversely
correlates with surface brightness.

However, there are two strong arguments against this notion.  First, it
implies the existence of ``naked'' O stars which we should see directly in HST images
of the nearest low surface brightness galaxies.  Instead, the color
magnitude diagrams of dwarf irregulars are typically deficient in high
mass stars \citep[e.g.][]{tgmf91,gmtf93,mtgf95,tolstoy96}.  While one
may argue that the high mass stars are completely hidden by dust this
would be inconsistent with them being naked.  Furthermore,  
high mass stars can be isolated in large numbers from
color-magnitude diagrams of higher surface-brightness dustier galaxies 
\citep[e.g.][]{atgsal05,g08,aamtvals08}.  Second, this
would require that dwarf LSB galaxies have an ISM that is more porous to ionizing
photons than more massive galaxies.  Dwarf galaxies with apparently
porous \HI\ distributions have been observed \citep[e.g.\
HoII;][]{pwbr92}.  However dwarf galaxies typically contain a higher
fraction of their mass in the ISM \citep[][H06]{vss95,sahs02} while
the lower mass densities of their disks suggests that that the ISM
distribution should be ``puffier''.  These factors should make it harder
for ionizing photons to escape, not easier.  Indeed, Using SINGG data
\citep{singg3} argue that \fesc\ increases with surface brightness.
While we can not directly rule out a variable \fesc\ as driving the
\Fhafuv\ correlations we believe these arguments are sufficiently strong
that a variable porosity scenario seems unlikely.

\subsection{Stochastic limitations}\label{s:stoc}

\citet{thilker07} and \citet{b07} use stochastic effects
to explain the often seen strong UV emission beyond the \Halpha\ truncation
radius of spiral galaxies \citep{mk01}.  They  
argue that the low \Fhafuv\ results from stars forming in events that are too low in
mass to have been likely to form O stars.  In other words, the IMF is
normal, but the star formation is too weak to produce the high
mass stars.  However, If the IMF is purely a statistical distribution of
the masses of stars formed in a single event, then by averaging over
many active star forming regions, one should recover the expected mean
\Fhafuv\ for the stellar population.  

We find that the galaxies in our sample are all bright enough that they
should have multiple O stars.  We calculate this as follows.  The
minimum SFR needed to be likely to see a single O star is given by
\begin{equation}
{\rm SFR_{min}} = \frac{\Mstaravg}{\langle \tau_{\rm O} \rangle} \frac{N_\star}{N_{\rm O}}
\end{equation}
where \Mstaravg\ is the average star mass, $\langle \tau_{\rm O}
\rangle$ is the typical lifetime of an O star, and $N_\star/N_{\rm O}$
is the ratio of the total number of stars and the stars formed above the
O star mass limit for the chosen IMF.  For a single power law IMF then
\begin{equation}
\Mstaravg = \left(\frac{\gamma+1}{\gamma+2}\right)
  \left(\frac{\Mupp^{\gamma+2} - \Mlow^{\gamma+2}}{\Mupp^{\gamma+1} - \Mlow^{\gamma+1}}\right)
\end{equation}
and 
\begin{equation}
\frac{N_\star}{N_{\rm O}} = \frac{\Mupp^{\gamma+1}-\Mlow^{\gamma+1}}{\Mupp^{\gamma+1}-\Most^{\gamma+1}}.
\end{equation}
Where we adopt $\Most = 20\,\Msun$ as the minimum mass of an O star.
For our fiducial IMF parameters (Sec~\ref{s:fmod}) then 
$\Mstaravg = 0.35\, \Msun$, $N_\star/N_{\rm O} = 1510$.
Table~\ref{t:mupmods} shows that for this IMF, and long duration star
formation, the median mass contributing to \LHa\ is $\Mstar = 54$ \Msun.
We adopt the main sequence lifetime of such a star as
$\langle \tau_{\rm O} \rangle = 4.0$ Myr \citep{bfbc93}, hence the SFR
must exceed ${\rm SFR_{min}} = 1.3\times10^{-4}\, \Msun\, {\rm
year}^{-1}$ for it to be likely to observe one O star.  This agrees very
well with the simulations of \citet{thilker07} showing that the
``stochastic limit'' for observing \Halpha\ emission is at $\sim
10^{-4}\, \Msun\, {\rm year}^{-1}$ for the same IMF parameters.  

For our fiducial stellar population, we calculate that ${\rm SFR_{min}}$ corresponds to
$M_{\rm FUV} = -8.29$ ABmag.  This is a
conservative estimate since shorter $\Delta t$ would result in a fainter
$M_{\rm FUV}$.  All the galaxies in our sample are significantly
brighter than this, and we calculate that they should have $N_{\rm O}
\gtrsim 16$ for continuous star formation with the fiducial IMF.  However,
we note that HIPASS J1321$-$31 has a pre-dust extinction correction
absolute magnitude of $M_{\rm FUV}' = -9.61$, the faintest in our
sample.  This implies that we are only likely to see three or four O stars in this
galaxy.  This is the only source that looks like a complete \Halpha\
non-detection in our sample.  With this few O stars likely to be present
at any one time, it could be that there is no \Halpha\ because the dust
absorption correction is overestimated and we caught the galaxy at ``a bad
time''.  Otherwise, our target galaxies are well beyond the limit where
statistical effects will bias \Fhafuv.

Stochastic effects may be even more important once plausible physical
constraints are put on the IMF.  Such an approach is adopted by
\citet{kw03} and \citet{wk05} who consider the clustered nature of star
formation, and impose the requirement that \Mupp\ of a given cluster is
statistically allowed for the mass of stars formed in the cluster
\Mclus.  We refer to this as the cluster mass constrained IMF scenario.
For an underlying $\gamma = -2.35$ (for the highest mass stars in a
broken power law IMF following Kroupa 2001, 2002) this approach results
in a strong correlation between \Mupp\ and \Mclus\ for low masses
($\Mupp \propto \Mclus^{0.67}$) that asymptotes to the true stellar
$\Mupp \approx 150$ \Msun\ for clusters having $\Mclus > 6800$ \Msun\
\citep{pwk07}.  For a power-law cluster mass function $\Mclus \propto
{\cal M}^{-2}$ the resultant IGIMF slope steepens to $\gamma \approx
-2.8$ for the highest mass stars \citep{kw03,wk05}.  The cluster mass
constrained IMF scenario predicts that traditional \LHa\ to SFR
conversion factors underestimate the true SFR, especially for SFR $<
10^{-2}\, \Msun\, {\rm year}^{-1}$, up to two orders of magnitude higher
than our estimate of ${\rm SFR_{min}}$. The scenario can also account
for the correlation between the luminosity of the brightest young star
cluster and the SFR observed in galaxies \citet{wkl04}. In a similar vein,
\citet{thilker07} also simulate the effects of requiring a certain mass
of stars to form in each cluster before high mass stars can form, and
find that such a prescription could result in an order of magnitude
increase in the stochastic limit ${\rm SFR_{min}}$. This could provide a
reasonable explanation for the low \Fhafuv\ values observed in LSB
galaxies and outer disks.  Indeed, \citet{pk08} show the cluster mass
constrained IMF scenario can produce \Halpha\ edges and the non-linear
relationship between the projected ISM density and \SHa\ observed within
galaxies.  Further implications of this scenario for our results are
discussed in Sec.~\ref{s:implic}.

While the cluster mass constrained IMF scenario is based on an IMF that
has an invariant form, the resultant IGIMF is variable and only approaches
this form for very high SFRs.  Thus this can not be considered a
constant IMF, but rather an explanation of what could cause the IGIMF to
vary.  Furthermore, while the scenario is appealing, the
physical basis for it is somewhat elusive.  It in essence it is
replacing a probabilistic interpretation of the IMF ``the formation of a
high mass stars in a low mass cluster is unlikely'' with a deterministic
interpretation ``the formation of a high mass stars in a low mass cluster is
impossible''.  The implication is that \Mupp\ is set by the available
gas supply.  However, the formalism is based on the final stellar mass of the
cluster, not the gas mass of the progenitor.  Forming or ``embedded''
clusters are found in molecular clouds having $M \geq 100\, \Msun$
\citep{ll03}, sufficient to form at least one star with our adopted
\Mupp, so it is not the ISM supply that is limiting \Mupp.  In reality
the molecular clouds fragment, and especially for low mass clusters, the
star formation efficiency is low and the clusters do not remain bound
\citep{ll03}.  So in effect, nature may produce stars in a way that
resembles the cluster mass constrained IMF.  In Sec.~\ref{s:scene} we
discuss an alternative scenario for a variable IMF based on the physics
of high mass star formation.

\subsection{Metallicity}\label{s:metals}

Metallicity affects \Fhafuv\ via line blanketing, which affects the
stellar atmosphere temperatures.  We calculated Starburst99 stellar
population models with the same parameters as our fiducial population
model but with metallicities of $Z = 0.004$ and 0.05 (1/5 and 2.5
times solar metallicity).  These metallicities were used for both the
isochrones and the spectra used to calculate the fluxes.  These models
result in $\Fhafuv = 14.0$ and 8.2 respectively.  Decreasing metallicity
increases \Fhafuv\ as the hotter stellar atmospheres results in
relatively more ionizing photons compared to the non-ionizing UV
continuum.  
The models cover from SMC to super solar metallicities and result in a
$\sim 0.25$ dex change in \Fhafuv, far smaller than the range of the
\Fhafuv\ variations.  Since LSB dwarf galaxies typically have lower
metallicities than normal galaxies, any trends with $\Sigma$ should act
to reduce the observed correlations, rather than contribute to them.
Therefore, we conclude that metallicity does not contribute to the
observed correlations of \Fhafuv\ with surface brightness.

These results hold for equilibrium models with our fiducial
IMF parameters.  Further exploration of models having different
parameters show that for very steep $\gamma \lesssim -3.3$ and $Z =
0.004$ \Fhafuv\ becomes smaller than for solar metallicity models.
This is shown in Fig.~\ref{f:mods}b discussed below.  As metallicity
decreases lower mass stars make a larger contribution to the FUV
emission, and for very steep $\gamma$ values this becomes more important
than the increased output of ionizing photons.

\subsection{IMF variations}\label{s:imf}

\begin{figure*}
\epsscale{1.0}
\plotone{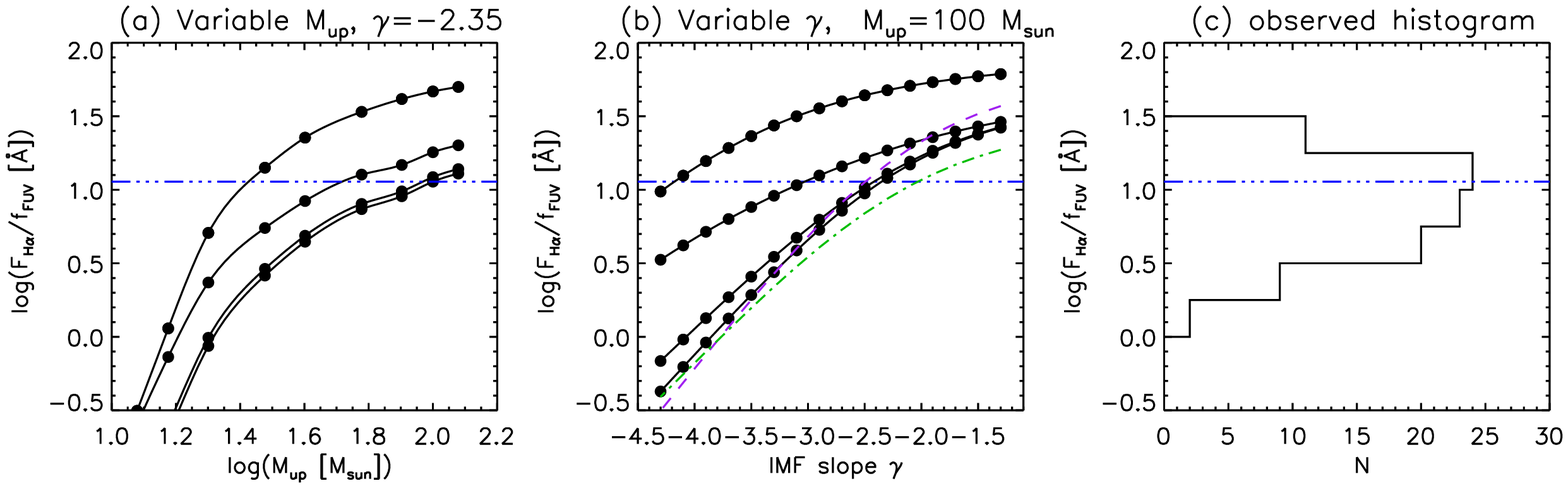}
\caption[]{Starburst 99 model predictions of \Fhafuv\ for stellar
  populations forming at a constant star formation rate are shown in
  panels (a) and (b).  Panel (c) shows the distribution of observed
  \Fhafuv\ values for comparison.  In panel (a) the IMF slope $\gamma$
  is held constant at the Salpeter (1955) value while \Mupp\ is varied.
  In panel (b) the upper mass limit is held constant at $\Mupp = 100\,
  \Msun$ while $\gamma$ is varied.  In these panels the dots indicate
  the \Mupp\ and $\gamma$ values used in the solar-metallicity ($Z =
  0.02$) models, while solid lines connect the models for star formation
  durations of 0, 10 Myr, 100 Myr, and 1 Gyr (from top to bottom).  
  The dashed (purple) and
  dot-dashed (green) lines in panel (b) indicate tracks with a
  metallicity $Z = 0.004$ and 0.05 respectively and a duration of 1
  Gyr.  The broken horizontal line in all panels shows the expected
  \Fhafuv\ for our fiducial stellar population model (eq.~\ref{e:grat}).  
  \label{f:mods}}
\end{figure*}

Figure~\ref{f:mods} show results of Starburst99 models designed to probe
the sensitivity of \Fhafuv\ to the IMF
parameters \Mupp\ and $\gamma$.  The models have constant
SFR over duration $\Delta t$ up to 1 Gyr, and (mostly) solar metallicity. We hold other
parameters fixed to the values of our fiducial stellar population model.
In panel a. \Mupp\ is varied and $\gamma = -2.35$ is held fixed, while
in panel b. $\gamma$ is varied and $\Mupp = 100$ \Msun is held fixed.
In this panel we also explore cases with $Z = 0.004$ and $Z = 0.05$ and $\Delta t = $1
Gyr in panel. Panel (c) shows the observed distribution of
\Fhafuv\ for comparison. We now consider whether the observed range of
\Fhafuv\ values can be modeled with long duration star formation by
varying \Mupp\ or $\gamma$.  The full sample (excluding upper limits)
spans $\Fhafuv = 1.5$ to 30\AA.  Since the extremes may be affected by
measurement errors, SFH variations, and stochastic effects we also
consider the tenth to ninetieth percentile range of \Fhafuv\ from 3 to
19\AA.

For a fixed (Salpeter) $\gamma = -2.35$ it is possible to reach the
lowest \Fhafuv\ with $\Mupp \sim 20$ \Msun, and the tenth percentile
with $\Mupp = 30$\Msun.  With equilibrium star formation models ($\Delta
t \gtrsim 300$ Myr) it is not possible to reach either the ninetieth
percentile or maximum \Fhafuv\ values.  The maximum $\Mupp = 120$ \Msun\
allowed by the Starburst99 models results in an equilibrium $\Fhafuv =
12.9$\AA.  The curves in Fig.~\ref{f:mods}a appear to be asymptotic
towards high \Mupp\ so it is unlikely that exploring higher \Mupp\
would help.  Lowering metallicity to $Z = 0.004$
brings the equilibrium $\Fhafuv = 14.0$ for $\Mupp = 100\, \Msun$ (Sec.~\ref{s:metals}) 
helping to alleviate the problem.  To achieve the highest observed
\Fhafuv\ values with solar metallicity and a Salpeter $\gamma$ requires
$\Delta t < 10$ Myr, shorter than the typical dynamical times of
galaxies (Sec.~\ref{s:sfh}), and thus rarely to be found in nature.

Varying $\gamma$ can explain a much larger range of the observed
\Fhafuv\ values.  Equilibrium star formation with $\gamma = -3.5$ can
account for the lowest \Fhafuv\ values, while the tenth percentile
corresponds to $\gamma = -3.3$.  The ninetieth percentile corresponds to
$\gamma = -1.9$, while the maximum \Fhafuv\ is slightly larger than the
equilibrium $\Fhafuv = 26$\AA\ for the shallowest $\gamma = -1.3$ we
considered. For $Z=0.004$ the maximum \Fhafuv\ corresponds to $\gamma
\approx -1.7$.

Having ruled out other possible explanations, we conclude that most of
the observed range of \Fhafuv\ values can most plausibly be explained
as arising from long duration star formation with an IMF that
varies at the upper end from galaxy to galaxy.  The lowest observed values can occur by lowering 
$\Mupp$ to $\sim 30$ \Msun, steepening $\gamma$ to $\sim -3.3$ or an
intermediate combination of the two.  It is harder to model the absolute highest
observed $\Fhafuv$ values with equilibrium star formation at solar
metallicity; lower metallicity can alleviate this problem.

\section{Discussion}\label{s:disc}

\subsection{Previous work}

Ours is not the first study to probe the nature of the IMF using
\Halpha\ and UV observations, nor the first to claim evidence of IMF
variations at the upper end.  \citet{bdd87} used UV observations from
the SCAP2000 experiment to explore the nature of the \Halpha\ to UV flux
ratio in a sample of bright spiral and irregular galaxies.  They found a
correlation in the ratio with morphological type, consistent to what we
find: early type spirals have higher \Fhafuv\ compared to late type
spirals and irregulars.  Technical issues (multiple instruments,
uncertain dust corrections) limited them from making strong claims about
the IMF.  They also noted a varying \fesc\ could also explain their
observed trend.  

Recently, \citet{hg08} deduced variations of the upper
end of the IMF using a very large ($\sim140$K galaxy) sample from the
Sloan Digital Sky Survey (SDSS) and employing an \Halpha\ equivalent
width versus optical color diagram, following the method pioneered by
\citet{k83}. They find that low luminosity galaxies have much lower
\ewha\ than expected for their colors and argue that this is due to the
upper end of the IMF varying systematically with galaxy mass.  The
variations are not monotonic with luminosity, but nevertheless similar
to our results: low luminosity galaxies have less massive stars than
higher luminosity galaxies.  Their result is solid, but since optical
bands are involved, much lower mass stars and longer timescales were
required in their modeling to rule out other explanations.

Our contribution to the field is to return to the basic method of
\citet{bdd87} with a larger sample and consistent dust absorption
corrections.  By using \Halpha\ and FUV measurements we isolate the
upper end of the IMF, and hence our constraints are more direct than
\Halpha\ -- optical color comparisons.  A new insight from our work is
that the surface brightness, or more likely surface mass
density, seems to be more important for driving the IMF
variations than luminosity or mass as suggested by \citet{hg08}.  Our
use of use of \HI\ selection and measurement of \Halpha\ and UV images
results in an extended range of optical surface
brightness compared to SDSS (M06), allowing this finding, while our
inclusive sample selection eliminates the possibility that our results
stem from a non-representative sample. 

\subsection{A scenario for the observed correlations}\label{s:scene}

The belief that the form of the IMF is universal is largely based on
observations of star clusters.  \citet{kroupa01} showed that apparent
variations in $\gamma$ observed in young star clusters could result
purely from stochastic effects due to the limited number of massive stars.
A universal IMF is then a logical conclusion if all stars form in star clusters as argued
by \citet{ll03}.  

Recent work suggests that this paradigm should be
re-examined.  Theory and simulations indicate that the highest mass stars ($>
10\, \Msun$) {\em require\/} a dense cluster environment to form
efficiently.  It is hard to form these stars by simple collapse and
fragmentation - the stars ``turn on'' and stop forming before high
masses can be assembled.  One way around this is by the process of
competitive accretion whereby young proto-stars ``steal'' material from
each others envelopes as well as accreting it from the dense ISM at the
bottom of a cluster's potential well \citep{bvb04}.  This allows rapid
assembly of massive stars but requires very dense environments, like
star clusters, and especially their dense centers \citep{bvb04}.  
While all stars may form in clusters, not all clusters are
alike. Clusters with a wide range of densities are known many of which
are unbound \citep{ll03}.  Ephemeral unbound clusters are not likely to
maintain a high enough density for long enough to have the collisions
necessary to form the most massive stars.

Pressure plays a key role in regulating the phases of the ISM
\citep{mo77,wmht03,br06}, and is the most likely physical property for
driving the observed \Fhafuv\ correlations.  While the neutral phases
(cold and warm) are in pressure equilibrium with each other, the
molecular ISM is self gravitating and supported by turbulence from star
formation feedback \citep{ds03}.  \citet{br06} show that the
ratio of molecular to neutral ISM in galaxies has a nearly linear
relationship with the mid-plane pressure, \Pext, expected from
hydrostatic equilibrium of a thin gas layer embedded in stellar disk
with a significantly larger scale height.  This leads to the expectation
that
\begin{equation}
\Pext \propto \Sigma_\star^{0.5}\Sigma_g(\sigma_g/\sqrt{h_\star}),
\end{equation} 
where $\Sigma_\star$ is the stellar mass density in the disk, $\Sigma_g$
is the ISM mass density, $\sigma_g$ is the ISM velocity dispersion, and
$h_\star$ is the stellar scale height.  The terms in parenthesis are not
expected to vary much within galaxies nor from galaxy to galaxy.
$\Sigma_\star$ varies strongly between galaxies and to first order is
$\propto \Sigma_R$.  The molecular to neutral mass
ratio in the ISM, and therefore the fraction of the ISM available for
star formation, should be highly dependent on $\Sigma_R$, because it
effectively is tracing pressure.  

\citet{ee97} and \citet{e08b} show that
tight bound clusters, and hence O stars, should preferentially form in a
high pressure environment while at lower pressures unbound and loose
clusters form (with less O stars).  While the pressure they are referring
to is that internal to the molecular clouds, \Pext\ provides a floor to
this pressure and thus should have some bearing on the final internal
pressures \citep{ds03}.  By this reasoning, \SR\ traces
\Pext\ which determines the likelihood that clusters are formed bound,
thus regulating the O/B ratio and therefore \Fhafuv.

The link between compact cluster formation and surface brightness is
well established observationally.  \citet{lr2000} and \citet{larsen04}
used ground based and HST images of spiral galaxies to show that
the fraction of $U$ band light in star clusters correlates with the
surface brightness of the host. Similarly, \citet{bhe02} show
that the
fraction of $U$ band light in star clusters scales almost linearly with
\SSFR.  \citet{m95} used Hubble Space Telescope NUV images to show that
there is a correlation between fraction of NUV light in the form of star
clusters and underlying surface brightness in starburst galaxies.  While
the correlation is weaker than that shown by \citet{larsen04}, the surface
brightness range is smaller, with the starbursts corresponding to the
high intensity, high cluster fraction extension of the sequence of Larsen.

Internal variations in \Fhafuv\ within a galaxy may also be related to
pressure variations.  For example, M83 shows a sharp decline in its
\Fhafuv\ ratio \citep{thilker05} corresponding to the \Halpha\
truncation radius identified by \citet{mk01}.  The FUV surface
brightness profile continues well beyond this radius, with no apparent
sign of a truncation.  Similar results are found in a larger sample by
\citet{b07} showing that \HII\ edges, generally are not seen in the
FUV.  \citet{mk01} show that the \HII\ edges corresponds well to where $\Sigma_g$ drops
below the critical density needed for disk self-gravity \citep{mk01}.
When a disk becomes gravitationally unstable it first excites low order
modes: bars and spiral arms.  These will induce spiral shocks which will
increase the ISM pressure and thus facilitate the formation of star
clusters and the efficient formation of O stars.  Thus 
star-formation edges may be a byproduct of the sharp-pressure increase
associated with spiral arms in massive unstable disks.

Our scenario is somewhat complementary to the more mathematically
developed cluster mass constrained IMF scenario of Kroupa and Weidner
discussed in Sec.~\ref{s:stoc}.  Indeed, they could be compatible to the
extent that our scenario offers a physical basis for small unbound
clusters having a truncated IMF.

\subsection{Implications of a variable top end of the IMF}\label{s:implic}

If the upper end of the IMF is not uniform then there are several
important astrophysical implications.  First, the star formation rate
(SFR) estimated for a galaxy depends on the tracer used.  While \Halpha\
and UV based SFR may agree for the bright galaxies, \Halpha\ based SFRs
will underestimate the contribution of LSB galaxies to the cosmic star
formation density of the universe.  Even if an alternate explanation of the \Fhafuv\ variations were
found  (e.g.\ \fesc\ variations), SFR estimates will still be under-estimated from
\Halpha\ fluxes in LSB galaxies.  Optically selected surveys of star formation will have
results that depend on their surface brightness selection limits.
Perhaps little notice has been paid to the problem of low \Fhafuv\
values 
because most star formation surveys have concentrated on normal and 
HSB galaxies
where the ratio is closer to normal.   UV based SFR estimates should be
more secure than \Halpha\ estimates.  However, there is no guarantee that
IMF variations stop at the O - B star divide.  For the cluster mass
constrained IMF scenario \citet{pwk07} show that
when the SFR is very low (SFR $< 10^{-3}\,  \Msun\, {\rm yr}^{-1}$) that
the IGIMF deviates from the canonical IMF down to \Mstar\ of just a few
\Msun\ and that standard conversion factors between \LHa\ and SFR 
could underestimate the SFR by orders
of magnitude in such cases.  However, at the low mass
end, below $\sim 0.8$ \Msun, the IMF appears to be constant
\citep{fgw99,wghfhgs02}.  More work needs to be done to determine where 
between the masses of O stars and the sub-solar range that the IMF
switches from being variable to constant. 

Second, a varying IMF also implies that the feedback of the young
stellar populations onto the ISM in terms of energy output and the
return of chemically enriched material will also vary, with LSB galaxies
having less feedback of energy and mass per mass in stars formed. The
metallicity of the returned ISM will also be affected.  LSB galaxies
will have a lower [O/Fe] ratio and higher [N/O] ratio than normal or HSB
galaxies.  While such abundance anomalies are observed in LSB dwarf
galaxies \citep[e.g.][]{tautvaisiene07} the abundance patterns are
usually explained by galactic winds that preferentially expel oxygen
rich ejecta in galaxies with weak gravitational potentials
\citep{p92,p93,dh94,mf99,db99}.  Undoubtedly such winds occur (e.g.\
NGC~1569: Waller 1991; NGC~1705: Meurer et al.\
1992)\nocite{w91,mfdc92}, but they would be harder to produce in a LSB
dwarf compared to a HSB blue compact dwarf.  Instead of removing excess
metal content in a galactic wind, LSB galaxies may just not produce the
metals.  Recently, \citet{kwk07} were able to model the functional form
of the mass-metallicity and mass yield relations of
\citep{tremonti04} using a cluster mass constrained IMF model
described above (Sec~\ref{s:stoc}) effectively demonstrating this
scenario is plausible. 

Measurements of the SFH of nearby galaxies from color
magnitude diagrams (CMDs) can also be misinterpreted.  In such an analysis,
one typically assumes that stars form with a Salpeter IMF fully
populated up to $\sim 100\, \Msun$.  If \Mupp\ is lower, or $\gamma$
steeper than assumed, then one could spuriously infer recently truncated
or declining SFHs.  Indeed many galaxies in the
nearby universe show very few field high mass stars.  These include LSB dwarf irregulars
\citep[e.g.][]{tgmf91,gmtf93,tolstoy96}, but also some blue compact
dwarfs \citep{greggio98,cannon03}.  The CMD results can then be
interpreted as a SFH that is ``gasping''
\citep{mtgf95} or being in a ``post-burst'' phase \citep{cannon03}.
Gasping could also result from a bias inherent in the star formation
scenario: stars are harder to identify in crowded clusters, hence there
is a bias against finding stars with life times shorter than the
dissolution time of clusters \citep{mtgf95}.  However, there are also
cases where the gasping deduced from CMDs is consistent with weak
\Halpha\ measurements \citep{tolstoy96,cole07}.  Claims based on CMD
analysis that star formation is ``contracting'' or progressively halting
from large radius to an inner starburst \citep[e.g.][]{skillman03} may
also be cast in doubt if there is a radial variation of the IMF.

\section{Summary}\label{s:summ}

We have looked at correlations of the integrated \Halpha\ to far UV flux
ratio \Fhafuv\ with other global parameters in a sample of galaxies that
well represents the full range of star formation properties of galaxies
in the local universe.  There are strong correlations of \Fhafuv\ with
optical surface brightness in both \Halpha\ and the $R$ band.  Weaker
but significant correlations are found with $R$ band luminosity,
rotational amplitude as well as morphology.  Thus the systematic
variations of \Fhafuv\ are part of the family of observational galaxy
scaling relations, such as the Hubble sequence, the Tully-Fisher
relation, the mass-metallicity relation, the luminosity-surface
brightness relation and the star formation law.  However, unlike other
scaling relations, the \Fhafuv\ correlations appear to be more closely
related to high mass star formation intensity or stellar mass density
than halo mass.

We examined a variety of plausible explanations for the root causes of
these correlations and ruled most of them out as follows.  1.\ While dust
correction decreases \Fhafuv, the effect is not large enough
to have spuriously created the correlations, and can not fix the problem
of low \Fhafuv\ values which exists even before dust correction. 2.\
Recent variations in the SFR would need to have had 
large amplitudes (factor $\gtrsim 10$) and short
duration ($\lesssim 100$ Myr) to account for the range of
\Fhafuv.  The normal undisturbed morphology of the majority of our
sample belie such extreme events in their recent histories.
Furthermore, variations in the SFR
can not account for the
correlations of \Fhafuv\ with both \SHa\ and \SR.  Invoking SFH to
explain the low \Fhafuv\ in LSB galaxies would require all LSB galaxies
to be experiencing coordinated gasps in their SFR which violates the
Copernican principle.
3. Low \Fhafuv\ values can occur if ionizing photons
escape.  However, one would then expect to see naked O stars in Hubble
Space Telescope images of 
the nearest LSB galaxies, and typically they are not found.
Furthermore,  LSB galaxies tend to have a higher gas fraction, and
should have thicker disks than normal galaxies which argues against them
having a higher \fesc.  4. Stochastic effects can not cause the observed trends since
we are measuring over entire galaxies, and they all are
sufficiently luminous in the UV, even without dust correction, that they
should contain multiple O stars.  The related cluster mass constrained
IMF scenario of \citet{kw03} and \citet{wk05} may explain the results,
but in effect it does not describe a constant IMF but a variable one.
5.\ Metallicity has only a fairly small effect on \Fhafuv\ with lower
metallicity acting to increase \Fhafuv\ for a normal IMF.  This is also
in the wrong sense since LSB galaxies tend to have low metallicity and
low \Fhafuv.

This leaves the most plausible explanation of the results to be a
varying upper end of
the IMF.  The
observed range \Fhafuv\  can be explained if the upper mass
limit \Mupp\ varies between 30 and 120 \Msun, or the IMF slope
$\gamma$ varies between --1.3 and --3.3.  There is not enough
information to decide between these options or rule out variations in
both parameters.  

We sketch a scenario that can explain the observed correlations as
arising from the pressure regulated formation of tightly bound star
clusters.  These are the sites of the formation of the highest mass
stars. A steeper or truncated IMF occurs when stars form in loose or
unbound clusters.  Regions of high pressure favor the formation of bound
clusters resulting in an increased O/B star ratio and thus higher
\Fhafuv.  $R$ band surface brightness traces pressures since it is the
established stellar populations that dominate the disk potential and
hence determines the mid-plane pressure at hydrostatic equilibrium.

There are several major implications of these results.  The SFR
estimated from \Halpha\ and FUV measurements are not necessarily the
same.  This holds even if the correlations result from a variable \fesc\
instead of IMF variations.  If the IMF is to blame, then the return of
energy, momentum, and metals to the ISM from high mass stars is
decreased in LSB galaxies.  This provides an
alternative driver for the mass metallicity relation other than galactic winds.  Allowing
IMF variations makes the interpretations of Color-Magnitude Diagrams
difficult: an observed lack of high mass stars may be due to a truncated
IMF rather than
being due to a recent decrease in the star formation rate.  Hence
interpretation of CMD results showing gasping/post-burst
SFHs or radial truncations of star formation are
cast in to doubt.

Further work is required to pin-down the nature of the \Fhafuv\
variations. We are currently examining internal \Fhafuv\ variations
within our sample galaxies in order to determine the extent to which the
IMF variations are local.  More observations of LSB galaxies are needed
to firmly rule out a variable \fesc\ as the cause of the variations.  If
LSB galaxies do have a high \fesc\ this would imply that either their
\HII\ regions no longer follow case B ionization, or that they have many
naked O stars.  Spectroscopic tests for either scenario should be
possible.  Careful CMD analyses, including UV measurements, of HSB and
LSB galaxies can be used to determine how far down the IMF the
variations may occur.  Comparison of abundance ratios with surface
brightness may provide constraints on the relative importance of a
truncated IMF and starburst winds for determining the abundance patterns
in dwarf galaxies.

\acknowledgments

Primary support for the work presented here was obtained through NASA
Galex Guest Investigator grant GALEXGI04-0105-0009 and NASA LTSA grant
NAG5-13083 to G.R.\ Meurer.  Additional support in the early phase of
the SINGG project came through grants NASA grants NAG5-8279
(Astrophysical Data Products program), HST-GO-08201, and HST-GO-08113 to
G.R.\ Meurer.  This work benefited from useful discussions with Rob
Kennicutt, Evan Skillman, Chris Flynn, and Guinnevere Kauffmann as well
as constructive comments from the annonymous referee.  Some of the first
analysis work for this project occurred while G.R.\ Meurer was visiting
Germany in September 2007; he thanks his hosts in Munich - Sandra Savaglio and Uta
Grothkopf, and in Freising - Ruth Lang and Armin M\"uller, for their
hospitality.

{\it Facilities:} \facility{GALEX}, \facility{CTIO:1.5m}, \facility{CTIO:0.9m}

\end{document}